\documentclass[11pt,a4paper]{article}
\pdfoutput=1
\usepackage{jheppub}

\usepackage{graphicx,bm}
\usepackage{amsmath,amssymb,mathrsfs,verbatim,subfigure}

\def\ofo{ { {}_2 \! F_1 }}

\newcommand{\p}{\partial}

\newcommand{\f}{\frac}
\def\ofo{ { {}_{2\!}F_1 }}
\title{Enhanced thermal photon and dilepton production in strongly coupled $\mathcal{N}=4$ SYM plasma in strong magnetic field}

\author{Kiminad A. Mamo}

\affiliation{Department of Physics, University of Illinois,
Chicago, IL 60607-7059, USA}

\emailAdd{kabebe2@uic.edu}
\abstract{We calculate the DC conductivity tensor of strongly coupled $\mathcal{N}=4$ super-Yang-Mills (SYM) plasma in a presence of a strong external magnetic field $B\gg T^2$ by using its gravity dual and employing both the RG flow approach and membrane paradigm which give the same results. We find that, since the magnetic field $B$ induces anisotropy in the plasma, different components of the DC conductivity tensor have different magnitudes depending on whether its components are in the direction of the magnetic field $B$. In particular, we find that a component of the DC conductivity tensor in the direction of the magnetic field $B$ increases linearly with $B$ while the other components (which are not in the direction of the magnetic field $B$) are independent of it. These results are consistent with the lattice computations of the DC conductivity tensor of the QCD plasma in an external magnetic field $B$. Using the DC conductivity tensor, we calculate the soft or low-frequency thermal photon and dilepton production rates of the strongly coupled $\mathcal{N}=4$ SYM plasma in the presence of the strong external magnetic field $B\gg T^2$. We find that the strong magnetic field $B$ enhances both the thermal photon and dilepton production rates of the strongly coupled $\mathcal{N}=4$ SYM plasma in a qualitative agreement with the experimentally observed enhancements at the heavy-ion collision experiments.

}


\arxivnumber{}

\begin{document}

\maketitle

\section{\label{sec:introduction}Introduction}
AdS/CFT correspondence since its inception in 1997 \cite{Maldacena:1997re,Gubser:1998bc,Witten:1998qj,Son:2002sd} has continued to give us invaluable insights into strongly coupled systems. For example, using the AdS/CFT correspondence, the shear viscosity to entropy density ratio of a strongly coupled quark-gluon plasma has been calculated \cite{Kovtun:2003wp,Buchel:2004di,Rebhan:2011vd,Mamo:2012sy} which turned out to be very small and close to the experimentally measured value at the heavy-ion collision experiments at $\mathbf{RHIC}$ and $\mathbf{LHC}$, in quite contrast to, the perturbative calculations at weak coupling which predicted a very large value \cite{Huot:2006ys}. The AdS/CFT correspondence computations have also given us numerous other qualitative insights into the heavy-ion collision experiments, see \cite{CasalderreySolana:2011us} for a review.

In this paper, we apply the AdS/CFT correspondence, to compute soft-thermal photon and dilepton production rates in strongly coupled $\mathcal{N}=4$ super-Yang-Mills (SYM) plasma in the presence of strong external magnetic field $B\gg T^2$ hoping to find qualitative insights into the quark-gluon plasma produced at $\mathbf{RHIC}$ and $\mathbf{LHC}$ which were recently found to contain a strong magnetic field background at the order of $B\sim 4m_{\pi}^2$ at $\mathbf{RHIC}$ \cite{Kharzeev:2007jp} and $B\sim 15m_{\pi}^2$ at $\mathbf{LHC}$ \cite{Skokov:2009qp}, produced during the early times of the non-central heavy-ion collisions. The effects of this strong magnetic field backgrounds on different signatures of the quark-gluon plasma has recently been explored in different contexts \cite{Fukushima:2008xe,Kharzeev:2007jp,Son:2004tq,Yee:2009vw,Rebhan:2009vc,Kharzeev:2010gd,Newman:2005hd,Landsteiner:2011iq,Basar:2012gh,Basar:2012bp,Bzdak:2012fr,Tuchin:2012mf}, see \cite{Kharzeev:2012ph} for a review.

Thermal photons and dileptons are defined as direct photons and dileptons produced from interactions other than decay process in the presence of thermal background or quark-gluon plasma (QGP), and cover the low-momentum $p_T<$2GeV \cite{Turbide:2007mi} and intermediate-mass 1GeV$\leq$ $M\leq$3.2 GeV \cite{Manninen:2010yf} region of the total direct photon and dilepton production spectrums, respectively, in the heavy-ion collision experiments.

The experimentally measured thermal photon and dilepton productions at $\mathbf{RHIC}$ \cite{Adare:2008ab,Adare:2009qk,Adler:2005ig} have shown significant enhancement in comparison to the thermal perturbative QCD \cite{Arnold:2001ms,Liu:2008eh,Fries:2005zh} and relativistic hydrodynamics \cite{Turbide:2007mi,Liu:2008eh,Fries:2005zh,Manninen:2010yf} predictions, and the enhancements increase in more non-central collisions \cite{Liu:2008eh,Fries:2005zh} where the magnetic field is expected to be stronger. In addition, the experimental measurements show that, the enhancement of the thermal dilepton production increases with the decreasing of its invariant mass \cite{Manninen:2010yf}. Thus, we hope to reproduce these and other features of the thermal photon and dilepton production rates from our AdS/CFT correspondence calculations in strong magnetic field $B$.
%Experimental measurement of the thermal photon production at RHIC \cite{Adare:2008ab} has been successfully modeled using perturbative QCD at zero temperature, %thermal perturbative QCD \cite{Arnold:2001ms}, and relativistic hydrodynamics \cite{Turbide:2007mi,Liu:2008eh,Manninen:2010yf}. However, the experimentally %measured thermal dilepton production has shown significant enhancement in comparison to the thermal perturbative QCD and hydrodynamics predictions %\cite{Turbide:2007mi,Liu:2008eh,Manninen:2010yf}. In particular, the measurements show that the enhancement of the thermal dilepton production increases with %the decreasing of the invariant mass of the dileptons \cite{Manninen:2010yf}. And, we hope to reproduce these and other features of the thermal photon and %dilepton production rates from our AdS/CFT correspondence calculations in a strong magnetic field $B$.

%The experimentally measured thermal photon and dilepton productions at RHIC \cite{Adare:2008ab} have shown significant enhancement in comparison to, and the %enhancements increase in more non-central collisions where the magnetic field is expected to be stronger. In addition, the experimental measurements show that, %the enhancement of the thermal dilepton production increases with decreasing of its mass \cite{Manninen:2010yf}. And, we hope to reproduce this and other %features of the thermal photon and dilepton production rates from our AdS/CFT correspondence calculations, in strong magnetic field.

Previous studies of the thermal photon and dilepton production rates at strong coupling without magnetic field include: $\mathcal{N}=4$ super-Yang-Mills plasma with zero \cite{CaronHuot:2006te} and non-zero chemical potential \cite{Jo:2010sg}; a strongly coupled plasma with flavor and with zero \cite{Mateos:2007yp} and non-zero baryon chemical potential \cite{Mas:2008jz}; finite 't Hooft coupling corrections \cite{Hassanain:2011ce, Hassanain:2012uj, Hassanain:2011fn}; prompt photon production rate \cite{Baier:2012ax, Steineder:2012si}; strongly coupled anisotropic plasma \cite{Patino:2012py, Rebhan:2011ke}.

The outline of this paper is as follows: In section \ref{sec:spectral}, we review the mathematical relationships that exist between spectral functions and production rates of thermal photons and dileptons.

In section \ref{sec:DC}, we use the holographic RG flow approach \cite{Iqbal:2008by}, which is a generalization of the membrane paradigm used in \cite{Kovtun:2003wp}, to calculate the transversal and longitudinal DC conductivities of the quark-gluon plasma both in the absence and presence of the strong magnetic field background. And, we find that the DC conductivity in the direction of the magnetic field $B$ increases linearly with $B$ in a qualitative agreement with the lattice results \cite{Buividovich:2010tn,Kalaydzhyan}.

In section \ref{sec:spectralp}, we compute the soft or low-frequency thermal photon spectral functions using the DC conductivities computed in section \ref{sec:DC} both in the absence and presence of the strong magnetic field background. In the presence of the strong magnetic field background, we compute the soft-thermal photon production rate when the momentum is parallel and perpendicular to the external magnetic field $B$, separately.

In section \ref{sec:spectrald}, we compute soft-thermal dilepton spectral functions using the DC conductivities computed in section \ref{sec:DC} both in the absence and presence of the strong magnetic field background. In the presence of the strong magnetic field background, we compute the soft-thermal dilepton production rate when the momentum is parallel and perpendicular to the external magnetic field $B$, separately.

In section \ref{sec:production rates}, we use the soft-thermal photon and dilepton spectral functions computed in section \ref{sec:spectralp} and section \ref{sec:spectrald} to evaluate the soft-thermal photon and dilepton production rates in the absence and presence of the strong magnetic field background, and show that the thermal photon and dilepton production rate in the presence of the strong magnetic field background is enhanced at $\mathbf{RHIC}$ and $\mathbf{LHC}$ energies. Specifically, we find that the production of the thermal dileptons increases with the decreasing of their invariant mass in a qualitative agreement with the experimental observation at $\mathbf{RHIC}$ \cite{Manninen:2010yf}.

%while the thermal photon production rate is not significantly affected by the presence of the strong magnetic field background at RHIC and LHC energies.

In Appendix \ref{Membrane paradigm}, we re-derive the DC conductivities using the membrane paradigm which gives the same result as the holographic RG flow approach used in the main text. In Appendix \ref{Equations of motion}, we write down the equations of motion explicitly both for $B=0$ and $B\gg T^2$ cases and show that unlike for the $B=0$ case, where the soft or low-frequency limit of the thermal photons and dileptons is found in the limit $\omega\ll T$, for the $B\gg T^2$ case the soft or low-frequency limit can be found in the limit $\omega\ll \sqrt{\sqrt{3}B}$.

\section{\label{sec:spectral}Spectral functions, and thermal photon and dilepton production rates}
Let's consider a field theory in thermal equilibrium, and
let the photon interaction with matter be of the
form $eJ_\mu A^\mu$, if $\Gamma_\gamma$ denotes the number of photons emitted per unit time
per unit volume, then the rate is given by \cite{CaronHuot:2006te}
\begin{equation}\label{eq:photon-rate-general1}
 \frac{d\Gamma_\gamma }{d^3k}=Q_{\gamma}\eta_{\mu\nu}\chi^{\mu\nu}(K)\Big|_{k^0=|\mathbf{k}|}\ ,
 \end{equation}
where $\chi^{\mu\nu}(K)$
is the spectral function, proportional to the imaginary part
of the retarded current-current correlation function
\begin{equation}
\chi^{\mu\nu}(K)=-2\,{\rm Im}\, C^{\mu\nu}(K) \ ,
\end{equation}
where $C^{\mu\nu}$ is the retarded two-point function of conserved current $J^\nu$
\begin{equation}
  C^{\mu\nu}(K) = -i\int\!d^4X e^{-iK\cdot X}
  \theta(t)\langle [J^\mu(X), J^\nu(0)]\rangle \ .
\label{eq:Wightman}
\end{equation}
And, $Q_{\gamma}=\frac{e^2}{16\pi^3|\mathbf{k}|}n_{b}(k^0)$ where $n_{b}(k^0)=1/(e^{\frac{k^0}{T}}-1)$ is the Bose-Einstein distribution function, $T$ is the thermal equilibrium temperature of the plasma, $\eta_{\mu\nu}$=diag(-+++) is the Minkowski metric, and $K$ is a null four-momentum vector with $k^0=|\mathbf{k}|=\omega$.

We can also re-write (\ref{eq:photon-rate-general1}) as
\begin{equation}\label{eq:photon-rate}
 \frac{d\Gamma_\gamma }{d\omega}=\overline{Q}_{\gamma}\chi^{\mu}_{\mu}(\omega)\ ,
 \end{equation}
where $\overline{Q}_{\gamma}=\frac{\alpha_{EM}T}{\pi}\frac{\frac{\omega}{T}}{e^{\frac{\omega}{T}}-1}$. And, for soft photons the spectral function $\chi^{\mu\nu}(\omega)$ is given in terms of the frequency independent conductivity (DC conductivity) $\sigma^{\mu\nu}$ as \cite{Policastro:2002se}
\begin{equation}\label{eq:spectral1}
\chi^{\mu\nu}(\omega)\cong2\omega\sigma^{\mu\nu}\ ,
 \end{equation}
for small $\omega$. Note that equation (\ref{eq:spectral1}) can be obtained by inverting the Kubo's formula for DC conductivity $\sigma^{\mu\nu}$\cite{Policastro:2002se}
\begin{equation}\label{kubo}
  \sigma^{\mu\nu}=\lim_{\omega\rightarrow0}\frac{1}{2\omega}\int dt d\textbf{x} e^{i\omega t}\langle[J^{\mu}(x), J^{\nu}(0)]\rangle=\lim_{\omega\rightarrow0}\frac{1}{2\omega}\chi^{\mu\nu}(\omega).
\end{equation}

If we also add to the above theory massive leptons which carry
only electric charge,
then the thermal system will also emit these leptons, produced
by virtual photon decay. Therefore, the same electromagnetic current-current correlation function,
evaluated for spacelike and timelike momenta $K^2=-M^2$, gives the dilepton production rate, \cite{CaronHuot:2006te}
\begin{equation}\label{eq:dilepton-rate-general1}
  \frac{d\Gamma_{\ell\bar\ell}}{d^4 K} = Q_{\ell\bar\ell}\chi^{\mu}_{\mu}(K)\,,
\end{equation}
where
  \begin{eqnarray}
 Q_{\ell\bar\ell} &=& \frac{1}{(2\pi)^4} \,  \nonumber
  \frac{e^2 \, e_\ell^2}{6\pi |K^2|^{5/2}} \
  \Theta(k^0)
  \Theta(-K^2{-}4m^2) \,\\
  &\times&
  [-K^2{-}{4m^2}]^{1/2} \,
  (-K^2{+}{2m^2}) \;
  n_{b}(k^0),
\end{eqnarray}
and, $e_\ell$ is the electric charge of the lepton,
$m$ is lepton mass, %and the correlator
%$C^<_{\mu\nu}(K)$ is evaluated at the timelike
%momentum of the emitted particle pair.
and $\Theta(x)$ denotes a unit step function.
Expressions (\ref{eq:photon-rate-general1}) and
(\ref{eq:dilepton-rate-general1}) for the production rates
are true to leading order in the electromagnetic couplings
$e$ and $e_\ell$,
but are valid non-perturbatively in all other interactions. And, for soft dileptons the spectral functions are given by the same equation as the soft photons (\ref{eq:spectral1}).

\section{\label{sec:DC}DC conductivities in $\mathcal{N}=4$ super-Yang-Mills plasma }
In this section, we calculate the DC conductivities of the $\mathcal{N}=4$ super-Yang-Mills plasma both in the absence $B=0$ and presence $B\gg T^2$ of the external magnetic field. Thus, for the case where the magnetic field background is present $B\gg T^2$, we calculate the DC conductivities separately when the momentum is parallel $k_z\parallel B_{z}$ and perpendicular $k_{x}\perp B_{z}$ to the magnetic field $B_{z}=B$.

\subsection{\label{sec:non-magnetic}DC conductivity for $B=0$}
The gravity dual of $\mathcal{N}=4$ super-Yang-Mills plasma at strong coupling and large $N_{c}$ limit is studied in an asymptotically $AdS_{5}$ metric \cite{CaronHuot:2006te}
\begin{eqnarray}\label{ads5}
ds^{2}&=&g_{\mu\nu}dx^{\mu}dx^{\nu}={\pi^2 T^2 R^2\over u}\left( -f(u) dt^2 + dx^2 + dy^2 +dz^2\right)+{R^2\over 4 f(u) u^2} du^2 \ ,
\end{eqnarray}
where $T = \frac{r_0}{\pi R^2}$ is the Hawking temperature which is conjectured to be the thermal equilibrium temperature of the plasma in section 2, $R^{4}=\lambda\ell_{s}^{4}$ is the radius of the $AdS_{5}$ spacetime, $\lambda=g^{2}_{YM}N_{c}$ is the 't Hooft coupling, $u = r_0^2/r^2$, $f(u)=1-u^2$, the horizon corresponds to $u=1$, the boundary to $u=0$, and the entropy density $s$ is given by
\begin{eqnarray}\label{s}
s=\frac{1}{4G_{5}}\sqrt{g_{xx}g_{yy}g_{zz}}=\frac{1}{2}\pi^2N_{c}^2T^3\ ,
\end{eqnarray}
where $G_{5}=\frac{\pi R^3}{2N_{c}^2}$ is Newton's constant. So, the energy density $\epsilon=\frac{3}{4}sT$ at infinite coupling $\lambda=\infty$ is
\begin{eqnarray}\label{si}
\epsilon=\frac{3}{8}\pi^2N_{c}^2T^4\ ,
\end{eqnarray}
while the zero coupling $\lambda=0$ result is
\begin{eqnarray}\label{s0}
\epsilon_{\lambda=0}=\frac{4}{3}\epsilon=\frac{1}{2}\pi^2N_{c}^2T^4\ .
\end{eqnarray}
In contrast, for the large-$N_{c}$ QCD plasma at zero coupling, see for example \cite{Czajka:2012gq}, we've
\begin{eqnarray}\label{qcd}
\epsilon_{QCD}=\frac{\pi^2}{60}(4N_{c}^2+7N_{f}N_{c})T^4\ ,
\end{eqnarray}
and, comparing (\ref{qcd}) and (\ref{s0}) for $N_{c}=3$ and $N_{f}=3$, we can infer that $\epsilon_{SYM}=2.73\epsilon_{QCD}$ at zero coupling and similar difference can be expected at strong coupling $1\ll\lambda\ll N_{c}$ limit. Therefore, we have to take this qualitative difference between QCD and $\mathcal{N}=4$ SYM plasma in consideration, whenever we try to compare the AdS/CFT correspondence computations in this paper with the heavy-ion collision experiments at $\mathbf{RHIC}$ and $\mathbf{LHC}$.

The gauge fluctuation $A_{\mu}$ is governed by the Maxwell's action
 \begin{equation}\label{action1}
 S=-\f{1}{4g^2_{5}}\int d^{d+1}x\sqrt{-g}F_{MN}F^{MN},
 \end{equation}
where $g^{2}_{5}=\frac{16\pi^{2}R}{N_{c}^2}$ \cite{CaronHuot:2006te}.

Choosing a gauge at which $A_{u}=0$ and choosing the wave to move in the $z$ direction only, i.e., $K = (\omega,0,0,k_{z})$, the equation of motion for the transversal component $A_{x}$ derived from the action (\ref{action1}) can be written as
\begin{eqnarray}\label{eom11}
\p_u\bigg(\f{1}{g^2_{5}}\sqrt{-g}g^{uu}g^{xx}A'_x\bigg)-\f{1}{g^2_{5}}\sqrt{-g} g^{xx}A_{x}(\omega^2 g^{tt}+k_{z}^2 g^{zz})&=&0. %\\
%\p_u\bigg[\f{\sqrt{-g}}{g^2_{5}} \f{g^{uu}g^{zz}A'_{z}}{(1-\frac{k_{z}^2}{\omega^2}\frac{g_{tt}}{g_{zz}})} %\bigg]+\f{\sqrt{-g}g^{tt}g^{zz}A_{z}\omega^2}{g^2_{5}}&=&0.
\end{eqnarray}

One also finds the current or the conjugate momentum to be
\begin{equation}\label{current}
 J^{x}=\frac{\partial\mathcal{L}}{\partial \partial_{u}A_{x}}=-\f{1}{g^2_{5}}\sqrt{-g}F^{ux}=-\f{1}{g^2_{5}}\sqrt{-g}g^{uu}g^{xx}A'_{x}.
\end{equation}
Then, using Ohm's law, defining the transversal %and longitudinal
frequency and momentum dependent (AC) conductivities at finite UV cut-off $u=\epsilon$ as $\sigma_{T}^{yy}(\epsilon,\omega,k_{z})=\sigma_{T}^{xx}(\epsilon,\omega,k_{z})=\f{J^x}{i\omega A_{x}}$, %and $\sigma^{L}=\sigma^{zz}=\f{J^z}{i\omega A_{z}}$,
one can derive the RG flow equation for the transversal AC conductivity $\sigma_{T}^{xx}(\epsilon,\omega,k_{z})$ using (\ref{eom11}) as \cite{Iqbal:2008by}
\begin{equation}\label{actransflow111}
    \partial_\epsilon\sigma_{T}^{xx} = \omega\sqrt{\frac{g_{uu}}{g_{tt}}}\left[\frac{(\sigma_{T}^{xx})^2}{\Sigma^{xx}(\epsilon)}-\Sigma^{xx}(\epsilon)\left(1+\frac{k_{z}^2g^{zz}}{\omega^2 g^{tt}}\right)\right]\ ,
\end{equation}
%while for the longitudinal AC conductivity $\sigma^{T}(\omega,k_{z},\epsilon)$ \cite{Iqbal:2008by}
%\begin{equation}\label{AC-long-flow}
%   \partial_\epsilon\sigma^{L} = -i\omega\sqrt{\frac{g_{uu}}{g_{tt}}}\left[\frac{(\sigma^{L})^2}{\Sigma^{L}(\epsilon)}\left(1-\frac{k_{z}^2g^{zz}}{\omega^2 %g^{tt}}\right)-\Sigma^{L}(\epsilon)\right]\ ,
%\end{equation}
where
\begin{eqnarray}\label{t1}
  %\Sigma^{L}(\epsilon) = \frac{1}{g^{2}_{5}}\sqrt{\frac{-g}{g_{uu}g_{tt}}}g^{zz}\ ,\\
  \Sigma^{xx}(u) = \frac{1}{g^{2}_{5}}\sqrt{\frac{g}{g_{uu}g_{tt}}}g^{xx}\ .
\end{eqnarray}
Since the right hand side of (\ref{actransflow111}) is divergent at the horizon $u=1$, requiring them to vanish there, due to the regularity condition at the horizon, we'll get, the momentum and frequency independent (DC) conductivities $\sigma_{T}^{yy}(\epsilon=1,\omega,k_{z})=\sigma_{T}^{xx}(\epsilon=1,\omega,k_{z})$ \cite{Iqbal:2008by}

\begin{equation}\label{txy}
\sigma_{T}^{xx}(\epsilon=1,\omega,k_{z})=\sigma_{T}^{yy}(\epsilon=1,\omega,k_{z})=\Sigma^{xx}(\epsilon=1)=\frac{1}{g^{2}_{5}}\sqrt{g_{xx}(1)g_{yy}(1)g_{zz}(1)}g^{xx}(1)=\frac{N_{c}^2T}{16\pi}=\sigma(1)\ .
\end{equation}

Similarly, if the wave is chosen to move in the $x$-direction, we'll have the transversal DC conductivities $\sigma_{T}^{zz}(\epsilon=1,\omega,k_{x})=\sigma_{T}^{yy}(\epsilon=1,\omega,k_{x})$
\begin{equation}\label{tzy}
\sigma_{T}^{zz}(\epsilon=1,\omega,k_{x})=\Sigma^{zz}(\epsilon=1)=\frac{1}{g^{2}_{5}}\sqrt{g_{xx}(1)g_{yy}(1)g_{zz}(1)}g^{zz}(1)=\frac{N_{c}^2T}{16\pi}=\sigma(1)\ .
\end{equation}

%Choosing a gauge at which $A_{u}=0$ and $A_{t}=0$ and choosing the wave to move in the $z$ direction only, i.e., $K = (\omega,0,0,k_{z})$, the equation of %motion for the longitudinal component $A_{z}$ derived from the action (\ref{action}), in momentum space, can be written as []
%\begin{eqnarray}\label{eom11L}
%\p_u\bigg[\f{\frac{1}{g^2_{5}}\sqrt{-g}g^{uu}g^{zz}A'_{z}}{1+\frac{k_{z}^2}{\omega^2}\frac{g^{zz}}{g^{tt}}} \bigg]+\f{1}{g^2_{5}}\sqrt{-g}g^{zz}A_{z}\omega^2 %g^{tt}=0.
%\end{eqnarray}
 We can also find the RG flow equation for the longitudinal component of the conductivity $\sigma_{L}^{zz}(\epsilon,\omega,k_{z})$, for example when the momentum is in the $z$ direction, from the equations of motion for the longitudinal component $A_{z}$ and the time component $A_{t}$ accompanied by the equation for the conservation of the current $J^{z}=-\f{1}{g^2_{5}}\sqrt{-g}F^{uz}$ and the Bianchi identity as \cite{Iqbal:2008by}
%One also finds the current or the conjugate momenta to be
%\begin{equation}\label{current}
% J^{z}=-\f{1}{g^2_{5}}\sqrt{-g}F^{uz}=-\f{1}{g^2_{5}}\sqrt{-g}g^{uu}g^{zz}A'_{z}.
%\end{equation}
%Then, using Ohm's law, defining the longitudinal
%conductivity at finite UV cut-off $u=\epsilon$ as $\sigma_{L}^{zz}(\epsilon,\omega,k_{z})=\f{J^z}{i\omega A_{z}}$, %and $\sigma^{L}=\sigma^{zz}=\f{J^z}{i\omega %A_{z}}$,
%one can derive the RG flow equation for the longitudinal momentum and frequency dependent (AC) conductivity $\sigma_{L}^{zz}(\epsilon,\omega,k_{z})$ using the %equation of motion (\ref{eom11L}) as \cite{Iqbal:2008by}
\begin{equation}\label{AC-long-flow}
  \partial_\epsilon\sigma_{L}^{zz} = \omega\sqrt{\frac{g_{uu}}{g_{tt}}}\left[\frac{(\sigma_{L}^{zz})^2}{\Sigma^{zz}(\epsilon)}\left(1+\frac{k_{z}^2g^{zz}}{\omega^2 g^{tt}}\right)-\Sigma^{zz}(\epsilon)\right]\ ,
\end{equation}
where
\begin{eqnarray}\label{t1}
  \Sigma^{zz}(\epsilon) = \frac{1}{g^{2}_{5}}\sqrt{\frac{g}{g_{uu}g_{tt}}}g^{zz} \ .
\end{eqnarray}
Since the right hand side of (\ref{AC-long-flow}) are divergent when the UV cut-off is at the horizon $\epsilon=1$, requiring them to vanish there, due to the regularity condition at the horizon, we'll get, the momentum and frequency independent (DC) longitudinal conductivity $\sigma_{L}^{zz}(\epsilon=1,\omega,k_{z})$ \cite{Iqbal:2008by}
\begin{equation}\label{l}
\sigma_{L}^{zz}(\epsilon=1,\omega,k_{z})=\Sigma^{L}(\epsilon=1)=\frac{1}{g^{2}_{5}}\sqrt{g_{xx}(1)g_{yy}(1)g_{zz}(1)}g^{zz}(1)=\frac{N_{c}^2T}{16\pi}=\sigma(1)\ .
\end{equation}

Similarly, if the wave is chosen to move in the $x$-direction, we get the longitudinal DC conductivity $\sigma_{L}^{xx}(\epsilon=1,\omega,k_{x})$
\begin{equation}\label{t2}
\sigma_{L}^{xx}(\epsilon=1,\omega,k_{x})=\Sigma^{xx}(1)=\frac{1}{g^{2}_{5}}\sqrt{g_{xx}(1)g_{yy}(1)g_{zz}(1)}g^{xx}(1)=\frac{N_{c}^2T}{16\pi}=\sigma(1)\ .
\end{equation}
Note that, throughout this paper, we work in the limit $k_{x}\approx \omega\ll T$ where the diffusion constant $D=0$.

\subsection{\label{sec:magnetic}DC conductivities for $B\gg T^2$}
Recently, a magnetic brane solution has been found in \cite{D'Hoker:2009mm,D'Hoker:2009bc} which interpolates between the $AdS_{5}$ spacetime (\ref{ads5}) in the UV or near the boundary and the $AdS_{3}\times T^2$ spacetime in the IR or near the horizon for $B\gg T^2$. Near the boundary, i.e., for $B\ll T^2$ the magnetic brane solution can be given as a perturbation series around the $AdS_{5}$ space (\ref{ads5}) in powers of $\frac{B}{T^2}$ \cite{D'Hoker:2009mm,D'Hoker:2009bc,Basar:2012gh} while the metric in the strong magnetic field $B\gg T^2$ regime is given by $AdS_{3}\times T^2$ metric \cite{D'Hoker:2009mm,D'Hoker:2009bc,Basar:2012gh}
\begin{equation}\label{ads3r1}
ds^{2}=g^{B}_{\mu\nu}dx^{\mu}dx^{\nu}=\frac{r^2}{(\frac{R}{\sqrt{3}})^2}\left(-f_{B}(r)dt^2+dz^2\right)+{(\frac{R}{\sqrt{3}})^2\over r^2f_{B}(r)}dr^2+\Big(\frac{R}{\sqrt{3}}\Big)^2(\sqrt{3}B dx^2+\sqrt{3}B dy^2)\ ,
\end{equation}
where $f_{B}(r)=1-\frac{r_{h}^2}{r^2}$, the horizon corresponds to $r=r_{h}$, the boundary to $r=\infty$, $R^{4}=\lambda\ell_{s}^{4}$ is the radius of the $AdS_{5}$ spacetime, and we can identify $\frac{R}{\sqrt{3}}$ as the radius of the $AdS_{3}$ spacetime. %that we get after compactifying the $x$ and $y$ %directions, which are transverse to the direction of the magnetic field $\sqrt{3}B=\sqrt{3}B^{z}=\sqrt{3}\epsilon^{zxy}F_{xy}$, into a 2-torus of radius %$\frac{R}{\sqrt{3}}$. Note that, since the radius of the 2-torus $\frac{R}{\sqrt{3}}$ has to be very large, so that the supergravity limit can be realized, we %may treat the 2-torus as if it's a 2-dimensional flat space.

Introducing $u = r_h^2/r^2$, we can re-write the metric (\ref{ads3r1}) in more convenient form as
\begin{equation}\label{ads3c}
ds^{2}=g^{B}_{\mu\nu}dx^{\mu}dx^{\nu}={\frac{4}{3}\pi^2T^2R^2\over u}\left(-f_{B}(u)dt^2+dz^2\right)+{R^2\over 12f_{B}(u)u^2}du^2+B\frac{R^2}{\sqrt{3}}(dx^2+dy^2)\ ,
\end{equation}
where $T = \frac{r_h}{\frac{2}{3}\pi R^2}$ is the Hawking temperature \cite{Basar:2012gh}, $\lambda=g^{2}_{YM}N_{c}$, $f_{B}(u)=1-u$, and the horizon corresponds to $u=1$. The entropy density $s_{B}$ is given by \cite{D'Hoker:2009mm,D'Hoker:2009bc}
\begin{eqnarray}\label{sb}
s_{B}=\frac{1}{4G_{5}}\sqrt{g^B_{xx}g^B_{yy}g^B_{zz}}=\frac{1}{3}N_{c}^2BT\ ,
\end{eqnarray}
where $G_{5}=\frac{\pi R^3}{2N_{c}^2}$ is Newton's constant. Comparing (\ref{sb}) and (\ref{s}), one can see that
\begin{eqnarray}\label{sb2}
s_{B}=\frac{2}{3\pi^2}\frac{B}{T^2}s=\frac{8}{3}bs\ ,
\end{eqnarray}
where we've defined the dimensionless quantity $b=\frac{B}{4\pi^2 T^2}$, and the ratio of the energy densities $\epsilon_{B}=\frac{3}{4}Ts_{B}$ and $\epsilon=\frac{3}{4}Ts$, at infinite coupling $\lambda=\infty$, will be
\begin{eqnarray}\label{esym1}
\frac{\epsilon_{B}}{\epsilon}=\frac{2}{3\pi^2}\frac{B}{T^2}\simeq0.07\frac{B}{T^2}\ ,
\end{eqnarray}
which can be compared to the zero coupling $\lambda=0$ result \cite{D'Hoker:2009mm,D'Hoker:2009bc}
\begin{eqnarray}\label{esym2}
\frac{\epsilon^{\lambda=0}_{B}}{\epsilon^{\lambda=0}}=\frac{\sqrt{3}}{2}\frac{3}{4}\frac{\epsilon_{B}}{\epsilon}\simeq0.05\frac{B}{T^2}\ .
\end{eqnarray}
So, for $\mathcal{N}=4$ super-Yang-Mills plasma, the ratio of the energy densities interpolates between $0.05\frac{B}{T^2}$ at zero coupling to $0.07\frac{B}{T^2}$ at infinite coupling.

In contrast, using the fact that for QCD plasma $\epsilon_{QCD}^B=\frac{B^2}{8\pi\alpha_{EM}}$ in the presence of the magnetic field $B$ at zero coupling, see for example \cite{Tuchin:2012mf}, we can infer that
\begin{eqnarray}\label{esym2a}
\frac{\epsilon_{QCD}^{B}}{\epsilon_{QCD}}\simeq(6.8\frac{B}{T^2})\times\frac{\epsilon_{SYM}^{B}}{\epsilon_{SYM}}=0.05\frac{(6.8\frac{B}{T^2})\times B}{T^2}\ .
\end{eqnarray}
Thus, one can see that equation \ref{esym2} and \ref{esym2a} are equivalent with the replacement of
\begin{equation}\label{replacement}
B\leftrightarrow(6.8\frac{B}{T^2})\times B.
\end{equation}
Therefore, whenever we compare the AdS/CFT correspondence computations in this paper with the heavy-ion collision experiments at RHIC and LHC, we have to use about $6.8\frac{B}{T^2}$ times stronger magnetic field than actually produced at those experiments, i.e., $B=B_{SYM}=(6.8\frac{B_{actual}}{T^2})\times B_{actual}$. Note that we are making the above conclusion based on an observation at weak coupling but we expect the same conclusion to hold in the strong coupling limit $1\ll\lambda\ll N_{c}$, at least qualitatively.

%In contrast, for zero coupling QCD, the ratio of the energy densities is given by \cite{Tuchin:2012mf}
%\begin{eqnarray}\label{eqcd}
%\frac{\epsilon^{QCD}_{B}}{\epsilon^{QCD}}=0.45\frac{B^2}{T^4}\ .
%\end{eqnarray}
%For example, taking $B=4m_{\pi}^2$ and $T=1.58m_{\pi}$ at RHIC, and taking the ratio of (\ref{eqcd}) and (\ref{esym2}), we can conclude that in order to see %significant effects on the signatures of $\mathcal{N}=4$ SYM plasma due to the magnetic field, we need about 6.4 times stronger magnetic field than we need to %see its significant effects in the QCD plasma. Similarly, taking $B=15m_{\pi}^2$ and $T=2.18m_{\pi}$ at LHC, and taking the ratio of (\ref{eqcd}) and %(\ref{esym}), we can conclude that in order to see significant effects on the signatures of $\mathcal{N}=4$ SYM plasma due to the magnetic field, we need about %12.4 times stronger magnetic field than we need to see its significant effects in the QCD plasma. Therefore, we should take this quantitative differences %between QCD and $\mathcal{N}=4$ SYM plasmas in consideration whenever we try to compare the AdS/CFT correspondence computations in this paper with the heavy-ion %collision experiments at RHIC and LHC.

%So, using (\ref{sb2}), $\omega_{B}=\int

%T_{B}dS_{B}$ can be re-written in terms %of $\omega=\int TdS$ as
%\begin{eqnarray}\label{sb3}
%\omega_{B}=\frac{3}{2\pi^2}\frac{B}{T^2}\omega=\frac{3}{2}\overline{b}\omega\ .
%\end{eqnarray}
The equation of motion and the RG flow equations for $B\gg T^2$ are still given by (\ref{eom11}) and (\ref{actransflow111}), respectively, but this time using the $AdS_{3}\times T^2$ metric $g^{B}_{\mu\nu}$ (\ref{ads3c}). So, if we take the momentum $k_{z}$ to be in the $z$-direction, which is parallel to the direction of the magnetic field $B=B_{z}\parallel k_{z}$, then the transversal DC conductivities $\sigma_{T}^{xx B_{\parallel}}(1)=\sigma_{T}^{yy B_{\parallel}}(1)$ will be
 \begin{equation}\label{t21}
\sigma_{T}^{xx B_{\parallel}}(1)=\frac{1}{g^{2}_{5}}\sqrt{g^{B}_{xx}(1)g^{B}_{yy}(1)g^{B}_{zz}(1)}g_{B}^{xx}(1)=\frac{2}{\sqrt{3}}\frac{N_{c}^2T}{16\pi}=\frac{2}{\sqrt{3}}\sigma(1)\ ,
\end{equation}
while the longitudinal DC conductivity $\sigma_{L}^{zz B_{\parallel}}(1)$ will be
\begin{equation}\label{t23}
\sigma_{L}^{zz B_{\parallel}}(1)=\frac{1}{g^{2}_{5}}\sqrt{g^{B}_{xx}(1)g^{B}_{yy}(1)g^{B}_{zz}(1)}g_{B}^{zz}(1)=\frac{1}{2}\frac{N_{c}^2B}{16\pi^3T}=2b\sigma(1)\ ,
\end{equation}
where we used $\sigma(1)=\frac{N_{c}^{2}T}{16\pi}$ to get the last line. Therefore, one can see that the DC conductivity $\sigma_{T}^{xx B_{\parallel}}(1)$ is independent of $B$ and has increased by a factor of $\frac{2}{\sqrt{3}}$ when the momentum is parallel to the magnetic field $B_{z}\parallel k_{z}$.

Similarly, if we take the momentum $k_{x}$ in the $x$-direction, which is perpendicular to the direction of the magnetic field $B=B_{z}\perp k_{x}$, then the transversal DC conductivities $\sigma_{T}^{yy B_{\perp}}(1)\neq \sigma_{T}^{zz B_{\perp}}(1)$ will be
\begin{equation}\label{t22}
\sigma_{T}^{yy B_{\perp}}(1)=\frac{1}{g^{2}_{5}}\sqrt{g^{B}_{xx}(1)g^{B}_{yy}(1)g^{B}_{zz}(1)}g_{B}^{yy}(1)=\frac{2}{\sqrt{3}}\frac{N_{c}^2T}{16\pi}=\frac{2}{\sqrt{3}}\sigma(1)\ ,
\end{equation}
and
\begin{equation}\label{t23}
\sigma_{T}^{zz B_{\perp}}(1)=\frac{1}{g^{2}_{5}}\sqrt{g^{B}_{xx}(1)g^{B}_{yy}(1)g^{B}_{zz}(1)}g_{B}^{zz}(1)=\frac{1}{2}\frac{N_{c}^2B}{16\pi^3T}=2b\sigma(1)\ ,
\end{equation}
while the longitudinal DC conductivity $\sigma_{L}^{xx B_{\perp}}(1)$ will be
\begin{equation}\label{t22}
\sigma_{L}^{xx B_{\perp}}(1)=\frac{1}{g^{2}_{5}}\sqrt{g^{B}_{xx}(1)g^{B}_{yy}(1)g^{B}_{zz}(1)}g_{B}^{xx}(1)=\frac{2}{\sqrt{3}}\frac{N_{c}^2T}{16\pi}=\frac{2}{\sqrt{3}}\sigma(1)\ .
\end{equation}
Note that $\sigma_{T}^{yy B_{\perp}}(1)$ is independent of $B$ and has increased by a factor of $\frac{2}{\sqrt{3}}$ while $\sigma_{T}^{zz B_{\perp}}(1)$ has increased linearly with $b=\frac{B}{4\pi^2 T^2}$. The fact that the DC conductivities $\sigma_{T}^{xx B_{\perp}}(1)=\sigma_{T}^{yy B_{\perp}}(1)$ are independent of the magnetic field $B=B_{z}$ while $\sigma_{T}^{zz B_{\perp}}(1)$ increases linearly with $B=B_{z}$ has already been observed in the lattice computations for $T=0$ (see Figure 3 of \cite{Buividovich:2010tn} and Figure 2 of \cite{Kalaydzhyan}, see also \cite{Nam:2012sg,Kerbikov:2012vp} which is consistent with our strong magnetic field or low temperature regime $T\ll\sqrt{B}$.

\section{\label{sec:spectralp}Spectral functions of thermal photons in $\mathcal{N}=4$ super-Yang-Mills plasma}
In this section, we'll compute the spectral functions of photons in the low-frequency limit, $\omega\ll T$ for $B=0$ or $\omega\ll \sqrt{B}$ for $B\gg T^2 $, using the DC conductivities calculated in section \ref{sec:DC}.
\subsection{\label{sec:non-magnetic}Spectral function for $B=0$}
Using (\ref{eq:spectral1}) and choosing the momentum of the photon to lie in the $z$-direction $K=(\omega,0,0,k_{z}=\omega)$, we find the transversal components of the spectral function $\chi^{xx}(\omega)$ and $\chi^{yy}(\omega)$ to be
\begin{eqnarray}
\chi^{xx}(\omega)=\chi^{yy}(\omega)&=&2\omega\sigma_{T}^{xx}(1)=2\omega\sigma(1)\ .
\end{eqnarray}
Then, we can calculate the trace of the spectral function $\chi^{\mu}_{\mu}(\omega)$ as
\begin{eqnarray}
\chi^{\mu}_{\mu}(\omega)&=&\chi^{t}_{t}(\omega)+\chi^{z}_{z}(\omega)+\chi^{x}_{x}(\omega)+\chi^{y}_{y}(\omega)=\chi^{x}_{x}(\omega)+\chi^{y}_{y}(\omega)=4\omega\sigma(1)\ ,
\end{eqnarray}
where we used the Ward identity $\frac{k_{z}^2}{\omega^2}\chi^{zz}=\chi^{tt}$, at light like momentum $k_{z}=\omega$, to eliminate the time and longitudinal components of the spectral function from its trace. The fact that only the transversal components of the spectral function contribute for the photon production rate has already been observed, for example in \cite{CaronHuot:2006te}.

Similarly, by making the momentum of the photon to lie in the $x$-direction $K=(\omega, k_{x}=\omega,0,0)$, one can find the transversal components of the spectral function $\chi^{yy}(\omega)$ and $\chi^{zz}(\omega)$ to be
\begin{eqnarray}
\chi^{yy}(\omega)=\chi^{zz}(\omega)&=&2\omega\sigma_{T}^{yy}(1)=2\omega\sigma(1)\ .
\end{eqnarray}
Hence, the trace of the spectral function $\chi^{\mu}_{\mu}(\omega)$ becomes
\begin{eqnarray}\label{low1}
\chi^{\mu}_{\mu}(\omega)&=&\chi^{y}_{y}(\omega)+\chi^{z}_{z}(\omega)=4\omega\sigma(1)\ .
\end{eqnarray}

For $B=0$, one can also calculate the trace of the spectral function $\chi^{\mu}_{\mu}(\omega)$ exactly at any frequency $\omega$, as it was first done in \cite{CaronHuot:2006te}, giving us an opportunity to compare our low-frequency result with the exact one. The exact result is \cite{CaronHuot:2006te}
\begin{equation}\label{exact}
\chi^{\mu}_{\mu}(\omega)=\chi^{y}_{y}(\omega)+\chi^{z}_{z}(\omega)=|_{2}F_{1}(1-(1+i)\frac{\omega}{4\pi T},1+(1-i)\frac{\omega}{4\pi T};1-i\frac{\omega}{2\pi T};-1)|^{-2}\omega\sigma(1)\ .
\end{equation}
So, using the identity $\ofo (1,1;1;-1)=\frac{1}{2}$ for Gauss's hypergeometric function $\ofo (a,b;c;z)$, it's clear that the exact result (\ref{exact}) reduces to our low-frequency result (\ref{low1}) in the $\frac{\omega}{T}\rightarrow 0$ limit.

\subsection{\label{sec:non-magnetic}Spectral functions for $B\gg T^2$}
Since, we have external magnetic field $B=B_{z}$ in the $z$-direction which creates anisotropy in our system, we'll carefully and separately study the spectral functions when the momentum is parallel and perpendicular to the direction of the magnetic field $B_{z}$.

\subsubsection{\label{sec:non-magnetic} $\mathbf{k_{z}}\parallel \mathbf{B_{z}}$}

Using (\ref{eq:spectral1}) and choosing the momentum of the photon to lie in the $z$-direction $K=(\omega,0,0,k_{z}=\omega)$, which is parallel to the magnetic field $B_{z}$, we find the transversal components of the spectral function $\chi^{xx B_{\parallel}}(\omega)$ and $\chi^{yy B_{\parallel}}(\omega)$ to be
\begin{eqnarray}
\chi^{xx B_{\parallel}}(\omega)=\chi^{yy B_{\parallel}}(\omega)&=&2\omega\sigma_{T}^{xx B_{\parallel}}(1)=\frac{4}{\sqrt{3}}\omega\sigma(1)\ ,
\end{eqnarray}
which means that the trace of the spectral function $\chi^{\mu B_{\parallel}}_{\mu}(\omega)$ becomes
\begin{equation}\label{spec1p}
\chi^{\mu B_{\parallel}}_{\mu}(\omega)=\chi^{x B_{\parallel}}_{x}(\omega)+\chi^{y B_{\parallel}}_{y}(\omega) =4\omega\sigma_{T}^{xx B_{\parallel}}(1)=\frac{8}{\sqrt{3}}\omega\sigma(1)\ .
\end{equation}

\subsubsection{\label{sec:non-magnetic} $\mathbf{k_{x}}\perp \mathbf{B_{z}}$}
Using (\ref{eq:spectral1}) and choosing the momentum of the photon to lie in the $x$-direction $K=(\omega,k_{x}=\omega,0,0)$, which is perpendicular to the magnetic field $B_{z}$, we find the transversal components of the spectral function $\chi^{yy B_{\perp}}(\omega)$ and $\chi^{zz B_{\perp}}(\omega)$ to be
\begin{eqnarray}
\chi^{yy B_{\perp}}(\omega)&=&2\omega\sigma_{T}^{B_{\perp},yy}(1)=\frac{4}{\sqrt{3}}\omega\sigma(1)\ ,
\end{eqnarray}
\begin{eqnarray}
\chi^{zz B_{\perp}}(\omega)&=&2\omega\sigma_{T}^{B_{\perp},zz}(1)=4b\omega\sigma(1)\ .
\end{eqnarray}
So, the trace of the spectral function $\chi^{\mu B_{\perp}}_{\mu}(\omega)$ becomes
\begin{equation}\label{spec1pr}
 \chi^{\mu B_{\perp}}_{\mu}(\omega)=\chi^{y B_{\perp}}_{y}(\omega)+\chi^{z B_{\perp}}_{z}(\omega) =2\omega\sigma_{T}^{B_{\perp},yy}(1)+2\omega\sigma_{T}^{B_{\perp},zz}(1)
=(\frac{4}{\sqrt{3}}+4b)\omega\sigma(1) \ .
\end{equation}

Finally, we would like to emphasize that our low-frequency limit results (\ref{spec1p}) and (\ref{spec1pr}) should be considered as a large magnetic field $B\gg T^2$ and low frequency $\omega\ll \sqrt{B}$ limits of a yet undetermined spectral functions at an arbitrary magnetic field $B$ and frequency $\omega$. Unfortunately, we couldn't find the exact spectral functions here since the exact bulk metric which interpolates between the $AdS_{3}\times T^2$ metric near the horizon for $B\gg T^2$, which we used in this paper, and the $AdS_{5}$ metric near the boundary is lacking \cite{D'Hoker:2009mm}.

\section{\label{sec:spectrald}Spectral functions of thermal dileptons in $\mathcal{N}=4$ super-Yang-Mills plasma}
In this section, we'll compute the spectral functions of soft or low-frequency dileptons, $\omega\ll T$ for $B=0$ or $\omega\ll \sqrt{B}$ for $B\gg T^2 $, using the DC conductivities calculated in section \ref{sec:DC}.

\subsection{\label{sec:non-magnetic}Spectral function for $B=0$}
Using (\ref{eq:spectral1}) and choosing the momentum of the dilepton to lie in the $z$-direction $K=(\omega,0,0,k_{z})$, we find the longitudinal and time components of the spectral function $\chi^{zz}(\omega,k_{z})$ and $\chi^{tt}(\omega,k_{z})$, respectively, to be
\begin{eqnarray}\label{spec1d}
 \chi^{zz}(\omega,k_{z})&=&2\omega\sigma_{L}^{zz}(1)=2\omega\sigma(1),
\end{eqnarray}
and
\begin{eqnarray}\label{spec1d}
 \chi^{tt}(\omega,k_{z})&=&\frac{k^2_{z}}{\omega^2}\chi^{zz}(\omega,k_{z})=2\frac{k^2_{z}}{\omega}\sigma(1),
\end{eqnarray}
where we used the Ward identity to find the time component of the spectral function $\chi^{tt}(\omega,k_{z})$ from the longitudinal one $\chi^{zz}(\omega,k_{z})$. Again, using (\ref{eq:spectral1}), we can find the transversal components of the spectral function for the dileptons to be
\begin{eqnarray}\label{spec1d}
 \chi^{xx}(\omega,k_{z})=\chi^{yy}(\omega,k_{z})&=&2\omega\sigma_{T}^{xx}(1)=2\omega\sigma(1).
\end{eqnarray}
So, the trace of the spectral function for the dileptons $\chi^{\mu}_{\mu}(\omega,k_{z})$ becomes
\begin{equation}\label{spec1dz}
\chi^{\mu}_{\mu}(\omega,k_{z})= \chi^{t}_{t}(\omega,k_{z})+\chi^{z}_{z}(\omega,k_{z})+\chi^{x}_{x}(\omega,k_{z})+\chi^{y}_{y}(\omega,k_{z})=-2\frac{k^2_{z}}{\omega}\sigma(1)+6\omega\sigma(1)\ .
\end{equation}

Similarly, when the momentum of the dileptons lies in the $x$-direction $K=(\omega,k_{x}=\omega,0,0)$, the trace of their spectral function $\chi^{\mu}_{\mu}(\omega,k_{z})$ becomes
\begin{equation}\label{spec1dx}
\chi^{\mu}_{\mu}(\omega,k_{x})= \chi^{t}_{t}(\omega,k_{x})+\chi^{x}_{x}(\omega,k_{x})+\chi^{y}_{y}(\omega,k_{x})+\chi^{z}_{z}(\omega,k_{x})=-2\frac{k^2_{x}}{\omega}\sigma(1)+6\omega\sigma(1)\ .
\end{equation}

\subsection{\label{sec:non-magnetic}Spectral functions for $B\gg T^2$}
Just like we did for the photons, we'll study the spectral functions of the dileptons when their momentum is parallel and perpendicular to the direction of the magnetic field $B_{z}$, separately.
\subsubsection{\label{sec:non-magnetic} $\mathbf{k_{z}}\parallel \mathbf{B_{z}}$}
Using (\ref{eq:spectral1}) and choosing the momentum of the dileptons to lie in the $z$-direction $K=(\omega,0,0,k_{z})$, which is parallel to the magnetic field $B_{z}$, we find the longitudinal and time components of the spectral function $\chi^{zz B_{\parallel}}(\omega,k_{z})$ and $\chi^{tt B_{\parallel}}(\omega,k_{z})$, respectively, to be
\begin{eqnarray}\label{spec1d}
 \chi^{zz B_{\parallel}}(\omega,k_{z})&=&2\omega\sigma_{L}^{zz B_{\parallel}}(1)=4b\omega\sigma(1),
\end{eqnarray}
and
\begin{eqnarray}\label{spec1d}
 \chi^{tt B_{\parallel}}(\omega,k_{z})&=&\frac{k^2_{z}}{\omega^2}\chi^{zz B_{\parallel}}(\omega,k_{z})=4b\frac{k^2_{z}}{\omega}\sigma(1).
\end{eqnarray}
We can also find the transversal components of the spectral function for the dileptons $\chi^{xx B_{\parallel}}(\omega,k_{z})$ and $\chi^{yy B_{\parallel}}(\omega,k_{z})$ to be
\begin{eqnarray}\label{spec1d}
 \chi^{xx B_{\parallel}}(\omega,k_{z})=\chi^{yy B_{\parallel}}(\omega,k_{z})&=&2\omega\sigma_{T}^{xx B_{\parallel}}(1)=\frac{4}{\sqrt{3}}\omega\sigma(1),
\end{eqnarray}
which means that the trace of their spectral function $\chi^{\mu B_{\parallel}}_{\mu}(\omega,k_{z})$ becomes
\begin{equation}\label{spec1dp}
\chi^{\mu B_{\parallel}}_{\mu}(\omega,k_{z})= \chi^{t B_{\parallel}}_{t}(\omega,k_{z})+\chi^{z B_{\parallel}}_{z}(\omega,k_{z})+\chi^{x B_{\parallel}}_{x}(\omega,k_{z})+\chi^{y B_{\parallel}}_{y}(\omega,k_{z})=-4b\frac{k^2_{z}}{\omega}\sigma(1)+4b\omega\sigma(1)+\frac{8}{\sqrt{3}}\omega\sigma(1)\ .
\end{equation}

\subsubsection{\label{sec:non-magnetic} $\mathbf{k_{x}}\perp \mathbf{B_{z}}$}
Again, using (\ref{eq:spectral1}) and choosing the momentum of the dileptons to lie in the $x$-direction $K=(\omega,k_{x},0,0)$, which is parallel to the magnetic field $B_{z}$, we find the longitudinal and time components of the spectral function $\chi^{xx B_{\perp}}(\omega,k_{x})$ and $\chi^{tt B_{\perp}}(\omega,k_{x})$, respectively, to be
\begin{eqnarray}\label{spec1d}
 \chi^{xx B_{\perp}}(\omega,k_{x})&=&2\omega\sigma_{L}^{xx B_{\perp}}(1)=\frac{4}{\sqrt{3}}\omega\sigma(1),
\end{eqnarray}
and
\begin{eqnarray}\label{spec1d}
 \chi^{tt B_{\perp}}(\omega,k_{x})&=&\frac{k^2_{x}}{\omega^2}\chi^{xx B_{\perp}}(\omega,k_{x})=\frac{4}{\sqrt{3}}\frac{k^2_{x}}{\omega}\sigma(1).
\end{eqnarray}
We can also find the transversal components of the spectral function for the dileptons $\chi^{yy B_{\perp}}(\omega,k_{x})$ and $\chi^{zz B_{\perp}}(\omega,k_{x})$, respectively, to be
\begin{eqnarray}\label{spec1d}
 \chi^{yy B_{\perp}}(\omega,k_{x})&=&2\omega\sigma_{T}^{yy B_{\perp}}(1)=\frac{4}{\sqrt{3}}\omega\sigma(1),
\end{eqnarray}
and
\begin{eqnarray}\label{spec1d}
 \chi^{zz B_{\perp}}(\omega,k_{x})&=&2\omega\sigma_{T}^{zz B_{\perp}}(1)=4b\omega\sigma(1).
\end{eqnarray}
So, the trace of the spectral function for the dileptons $\chi^{\mu B_{\perp}}_{\mu}(\omega,k_{x})$ becomes
\begin{equation}\label{spec1dpr}
\chi^{\mu B_{\perp}}_{\mu}(\omega,k_{x})= \chi^{t B_{\perp}}_{t}(\omega,k_{x})+\chi^{x B_{\perp}}_{x}(\omega,k_{x})+\chi^{y B_{\perp}}_{y}(\omega,k_{x})+\chi^{z B_{\perp}}_{z}(\omega,k_{x})=-\frac{4}{\sqrt{3}}\frac{k^2_{z}}{\omega}\sigma(1)+\frac{8}{\sqrt{3}}\omega\sigma(1)+4b\omega\sigma(1)\ .
\end{equation}

\section{\label{sec:production rates}Thermal photon and dilepton production rates}
In this section, we'll write down the photon and dilepton production rates (\ref{eq:photon-rate}) and (\ref{eq:dilepton-rate-general1}) using the trace of the spectral functions computed in section \ref{sec:spectralp} and \ref{sec:spectrald}, respectively.

\begin{figure}[ht]
\begin{center}
\begin{tabular}{cc}
\includegraphics[scale=0.6]{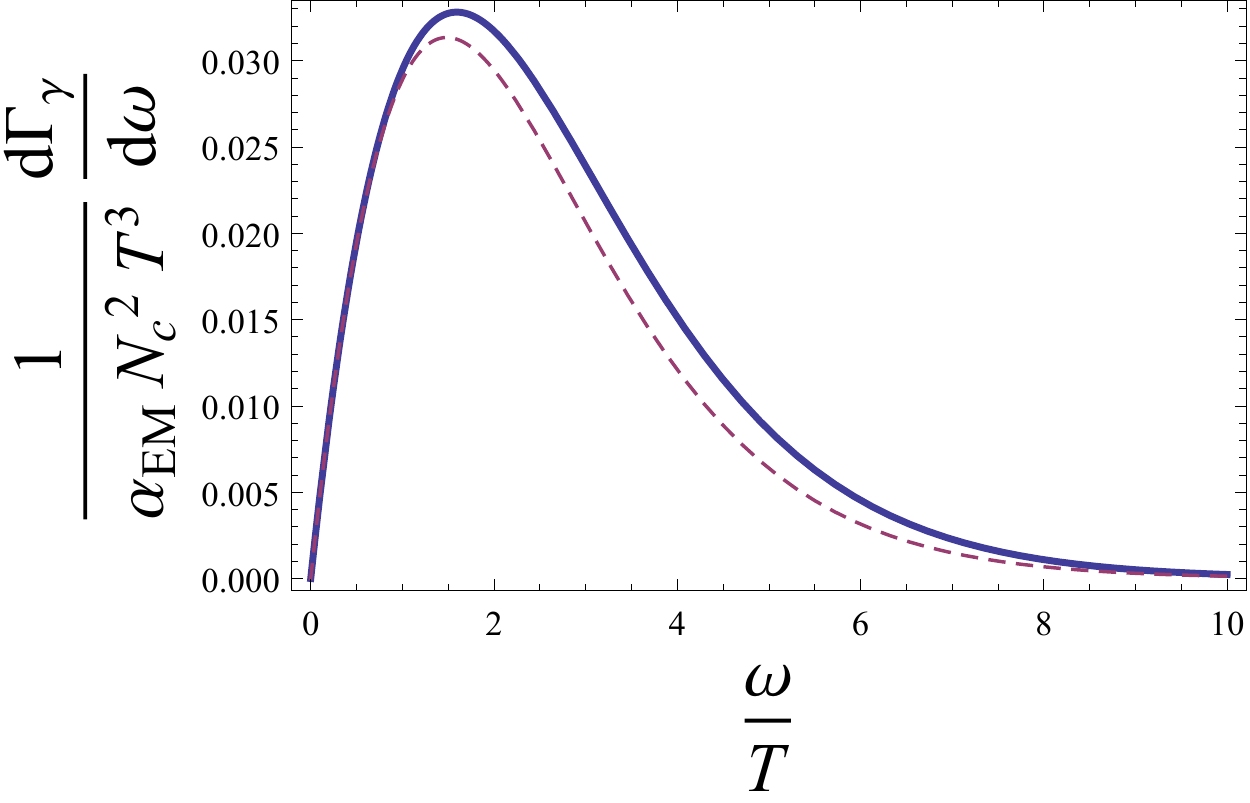}
%\put(-20,45){$a/T$}
%\put(-100,71){$\cb_4/T^4$}
%\put(-100,12){$\cf_4/T^4$}
&
\includegraphics[scale=0.6]{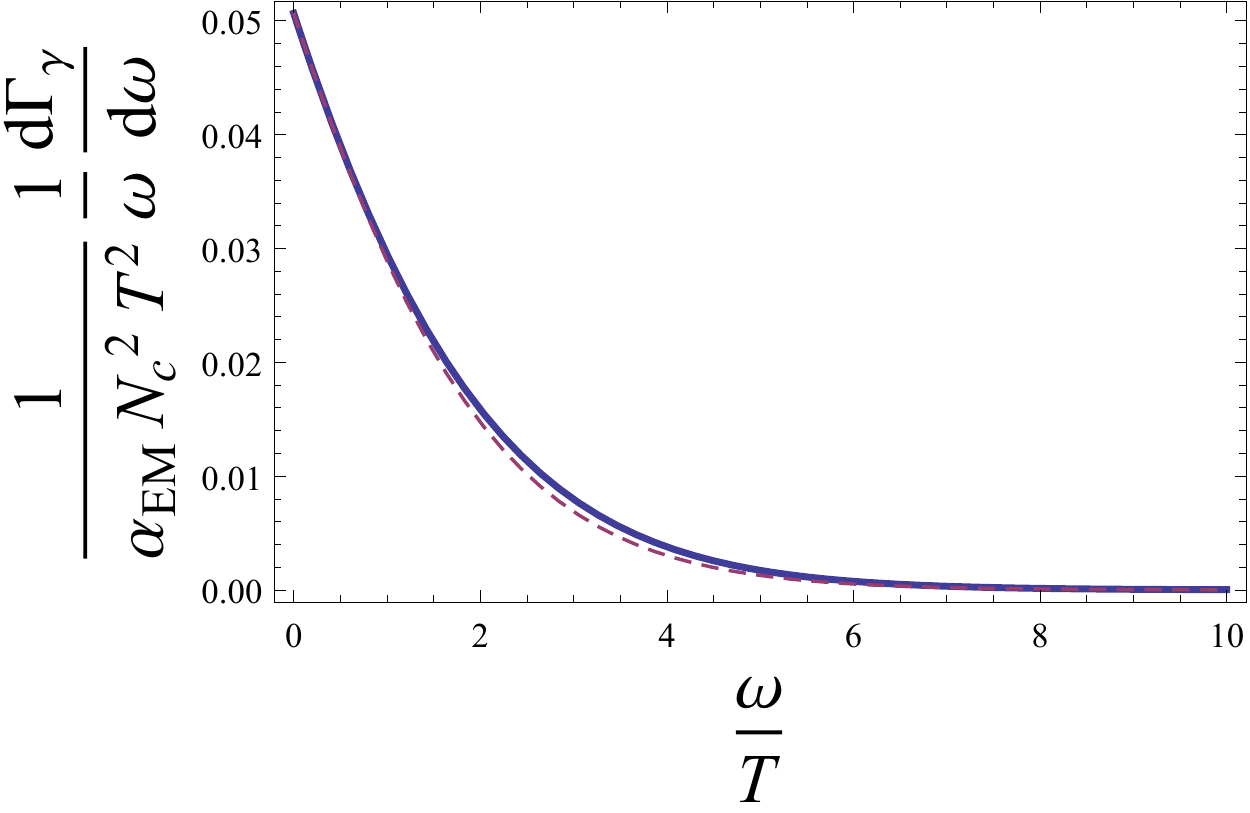}
%\put(-20,42){$T/a$}
%\put(-107,80){$\cb_4/a^4$}
%\put(-107,12){$\cf_4/a^4$}

\\
(a) & (b)
\end{tabular}
\caption{Thermal photon production at any frequency (\ref{any}) [\textit{solid lines}] and at low-frequency (\ref{low}) [\textit{dashed lines}] for $B=0$.
\label{low-exact}}
\end{center}
\end{figure}

The photon production rates for $B=0$ are found from (\ref{eq:photon-rate}) using the trace of the spectral functions (\ref{exact}) and (\ref{low1}), therefore, they are given by
\begin{equation}\label{any}
\frac{d\Gamma^{B=0}_\gamma }{d\omega}=|_{2}F_{1}(1-(1+i)\frac{\omega}{4\pi T},1+(1-i)\frac{\omega}{4\pi T};1-i\frac{\omega}{2\pi T};-1)|^{-2}\overline{Q}_{\gamma}\omega\sigma(1),
\end{equation}
for any frequency $\omega$, and
\begin{equation}\label{low}
\frac{d\Gamma^{B=0}_\gamma }{d\omega}\Big|_{\omega\ll T}=4\overline{Q}_{\gamma}\omega\sigma(1),
\end{equation}
for a small frequency $\omega\ll T$. We've compared the low-frequency result (\ref{low}) and the exact result (\ref{any}) in Fig.~\ref{low-exact}. Note that, in Fig.~\ref{low-exact}, we've multiplied both (\ref{low}) and (\ref{any}) by a factor of 2 in order to find the total thermal photon production rate in the $x$ and $z$ directions. Also, note that $\overline{Q}_{\gamma}\omega\sigma(1)=\frac{\alpha_{EM} N_{c}^2T^3}{16\pi^2}\frac{(\frac{\omega}{T})^2}{e^{\frac{\omega}{T}}-1}$.

\begin{figure}[ht]
\begin{center}
\includegraphics[scale=0.6]{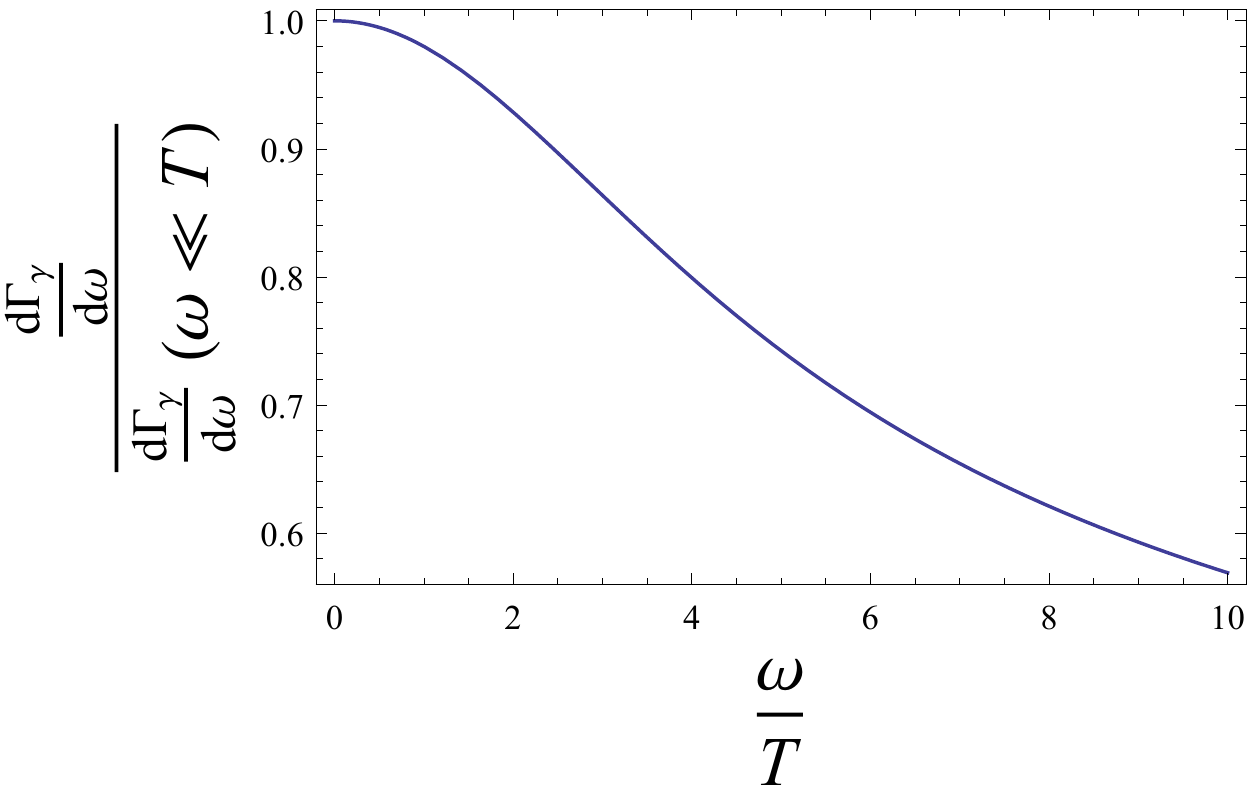}

\caption{The ratio of thermal photon production at any frequency (\ref{any}) and at low-frequency (\ref{low}) for $B=0$.
\label{ratio}}
\end{center}
\end{figure}

In the presence of strong external magnetic field $B_{z}=B\gg T^2$, for the photons with momentum $k_{z}=\omega$ parallel to the direction of the magnetic field $B_{z}$, the photon production rate (\ref{eq:photon-rate}) using the trace of the spectral function (\ref{spec1p}) becomes
\begin{eqnarray}\label{eq:photon-rate-general}
\frac{d\Gamma^{B_{\parallel}}_\gamma }{d\omega}\Big|_{\omega\ll \sqrt{B}}=\frac{8}{\sqrt{3}}\overline{Q}_{\gamma}\omega\sigma(1),
\end{eqnarray}
while for the photons with momentum $k_{x}=\omega$ perpendicular to the magnetic field $B_{z}$, the photon production rate (\ref{eq:photon-rate}) using the trace of the spectral function (\ref{spec1pr}) becomes
\begin{eqnarray}\label{eq:photon-rate-general}
 \frac{d\Gamma^{B_{\perp}}_\gamma }{d\omega}\Big|_{\omega\ll \sqrt{B}}=(\frac{4}{\sqrt{3}}+4b)\overline{Q}_{\gamma}\omega\sigma(1).
\end{eqnarray}

\begin{figure}[ht]
\begin{center}
\begin{tabular}{cc}
\includegraphics[scale=0.6]{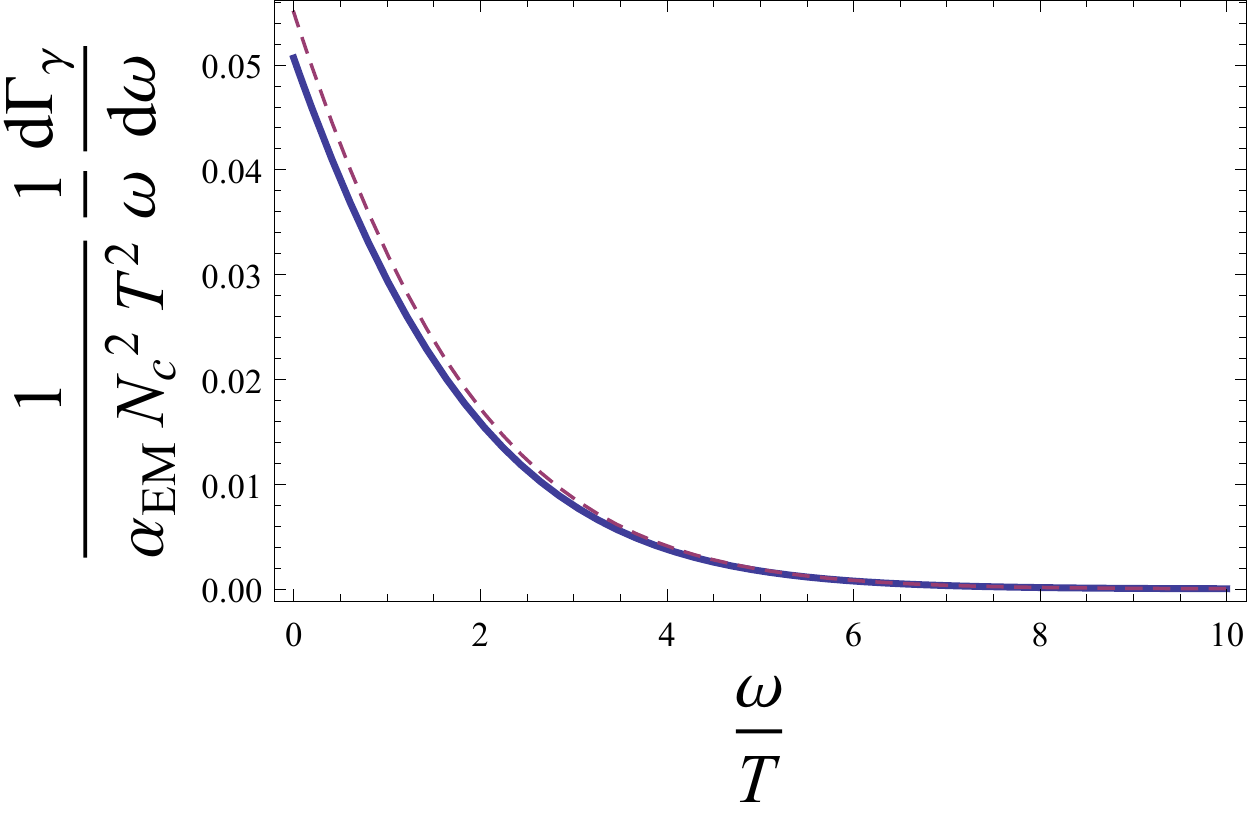}
%\put(-20,45){$a/T$}
%\put(-100,71){$\cb_4/T^4$}
%\put(-100,12){$\cf_4/T^4$}
&
\includegraphics[scale=0.6]{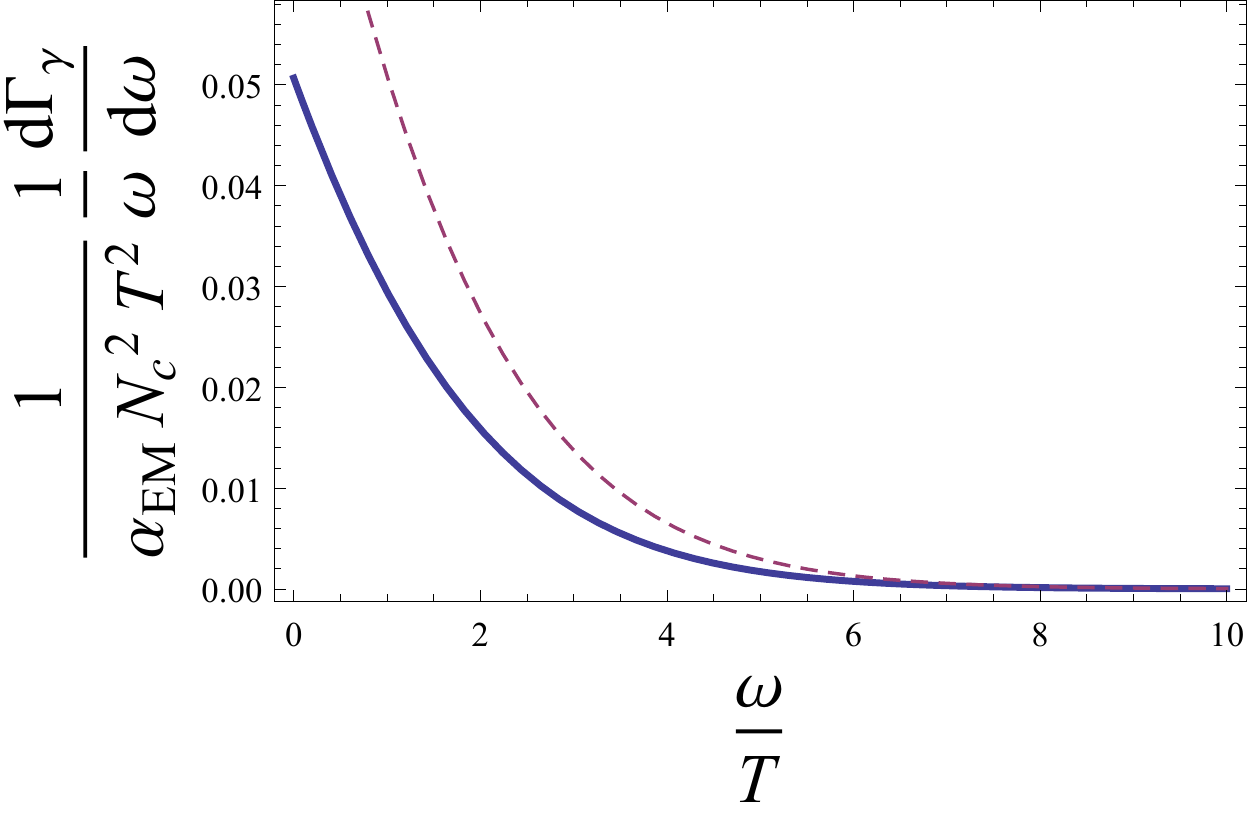}
%\put(-20,42){$T/a$}
%\put(-107,80){$\cb_4/a^4$}
%\put(-107,12){$\cf_4/a^4$}

\\
(\textbf{a}) & (\textbf{b})
\end{tabular}
\caption{Thermal photon production in the absence of the magnetic field $B=0$ (\ref{total1}) [\textit{solid lines}] and in the presence of the strong magnetic field $B\gg T^2$ (\ref{total2}) [\textit{dashed lines}]. In (\textbf{a}) we used (\ref{replacement}) to get $B=B_{SYM}=(6.8\frac{B_{RHIC}}{T^2})\times B_{RHIC}=10.90\times4m_{\pi}^2$, $T=1.58m_{\pi}$ and $b=\frac{1}{4\pi^2}\frac{B}{T^2}=0.44$ at \textbf{RHIC}. In (\textbf{b}) we used (\ref{replacement}) to get $B=B_{SYM}=(6.8\frac{B_{LHC}}{T^2})\times B_{LHC}=21.46\times15m_{\pi}^2$, $T=2.18m_{\pi}$ and $b=\frac{1}{4\pi^2}\frac{B}{T^2}=1.72$ at \textbf{LHC}.
\label{pbrl2}
}
\end{center}
\end{figure}

Finally, the total thermal photon production rates $\frac{d\Gamma^{B=0}_{\gamma(Total)}}{d\omega}$ and $\frac{d\Gamma^{B}_{\gamma(Total)} }{d\omega}$ become
\begin{equation}\label{total1}
 \frac{d\Gamma^{B=0}_{\gamma(Total)}}{d\omega}\Big|_{\omega\ll T}=2\frac{d\Gamma^{B=0}_\gamma }{d\omega}\Big|_{\omega\ll T}=8\overline{Q}_{\gamma}\omega\sigma(1),
\end{equation}
and
\begin{equation}\label{total2}
 \frac{d\Gamma^{B}_{\gamma(Total)} }{d\omega}\Big|_{\omega\ll \sqrt{B}}=\frac{d\Gamma^{B_{\parallel}}_\gamma }{d\omega}\Big|_{\omega\ll \sqrt{B}}+ \frac{d\Gamma^{B_{\perp}}_\gamma }{d\omega}\Big|_{\omega\ll \sqrt{B}}=(\frac{12}{\sqrt{3}}+4b)\overline{Q}_{\gamma}\omega\sigma(1).
\end{equation}
We have plotted the total thermal photon production rates (\ref{total1}) and (\ref{total2}) together in Fig.~\ref{pbrl2}. Note also that $\overline{Q}_{\gamma}\omega\sigma(1)=\frac{\alpha_{EM} N_{c}^2T^3}{16\pi^2}\frac{(\frac{\omega}{T})^2}{e^{\frac{\omega}{T}}-1}$. %Fig.~\ref{pbrl} and
%\begin{figure}[ht]
%\begin{center}
%\begin{tabular}{cc}
%\includegraphics[scale=0.6]{pr.pdf}
%%\put(-20,45){$a/T$}
%%\put(-100,71){$\cb_4/T^4$}
%%\put(-100,12){$\cf_4/T^4$}
%&
%\includegraphics[scale=0.6]{pl.pdf}
%\put(-20,42){$T/a$}
%\put(-107,80){$\cb_4/a^4$}
%\put(-107,12){$\cf_4/a^4$}

%\\
%(\textbf{a}) & (\textbf{b})
%\end{tabular}
%\caption{Thermal photon production in the absence of the magnetic field $B=0$ (\ref{total1}) [\textit{solid lines}] and in the presence of the strong magnetic %field $B\gg T^2$ (\ref{total2}) [\textit{dashed lines}]. In (\textbf{a}) we used $B=4m_{\pi}^2$, $T=1.58m_{\pi}$ and $b=\frac{1}{4\pi^2}\frac{B}{T^2}=0.04$ at %\textbf{RHIC}. In (\textbf{b}) we used $B=15m_{\pi}^2$, $T=2.18m_{\pi}$ and $b=\frac{1}{4\pi^2}\frac{B}{T^2}=0.08$ at \textbf{LHC}.
%\label{pbrl}
%}
%\end{center}
%\end{figure}

%The enhancement of the total photon production rate in the presence of the strong magnetic field $B$ can be quantified by taking the ratio
%\begin{equation}
%  x_{\gamma}=\frac{\frac{d\Gamma^{B}_{\gamma(Total)} }{d\omega}\Big|_{\omega\ll \sqrt{B}}}{\frac{d\Gamma^{B=0}_{\gamma(Total)}}{d\omega}\Big|_{\omega\ll %T}}=1.5(1+\frac{b}{\sqrt{3}}).
%\end{equation}

Similarly, the thermal dilepton production rate for $B=0$ is found from (\ref{eq:dilepton-rate-general1}) using the trace of the spectral function (\ref{spec1dz}), thus, it's given by
\begin{eqnarray}
\frac{d\Gamma^{B=0}_{\ell\bar\ell}}{d^4 K}\Big|_{\omega\ll T}=(-2\frac{k^2_{z}}{\omega^2}+6)Q_{\ell\bar\ell}\omega\sigma(1).
\end{eqnarray}
Therefore, in the presence of strong external magnetic field $B=B_{z}\gg T^2$, for the dileptons with momentum $k_{z}$ which is parallel to the direction of the magnetic field $B_{z}$, the thermal dilepton production rate (\ref{eq:dilepton-rate-general1}) using the trace of the spectral function (\ref{spec1dp}) becomes
\begin{eqnarray}
  \frac{d\Gamma^{B_{\parallel}}_{\ell\bar\ell}}{d^4 K}\Big|_{\omega\ll \sqrt{B}}= (-4b\frac{k^2_{z}}{\omega^2}+4b+\frac{8}{\sqrt{3}})Q_{\ell\bar\ell}\omega\sigma(1),
\end{eqnarray}
while for the dileptons with momentum $k_{x}$ perpendicular to the direction of the magnetic field $B_{z}$, the dilepton production rate (\ref{eq:dilepton-rate-general1}), using the trace of the spectral function (\ref{spec1dpr}), becomes
\begin{eqnarray}
 \frac{d\Gamma^{B_{\perp}}_{\ell\bar\ell}}{d^4 K}\Big|_{\omega\ll \sqrt{B}} =(-\frac{4}{\sqrt{3}}\frac{k^2_{x}}{\omega^2}+\frac{8}{\sqrt{3}}+4 b)Q_{\ell\bar\ell}\omega\sigma(1).
\end{eqnarray}

\begin{figure}[ht]
\begin{center}
\begin{tabular}{cc}
\includegraphics[scale=0.6]{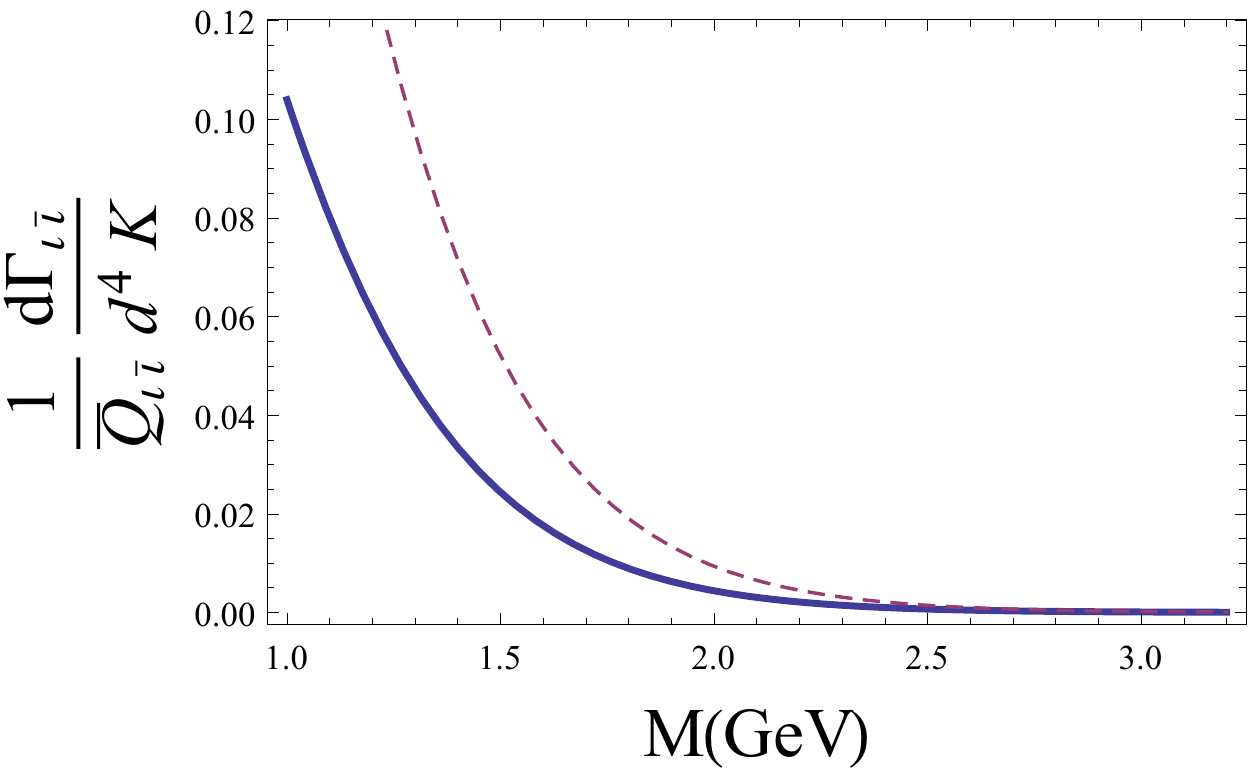}
%\put(-20,45){$a/T$}
%\put(-100,71){$\cb_4/T^4$}
%\put(-100,12){$\cf_4/T^4$}
&
\includegraphics[scale=0.6]{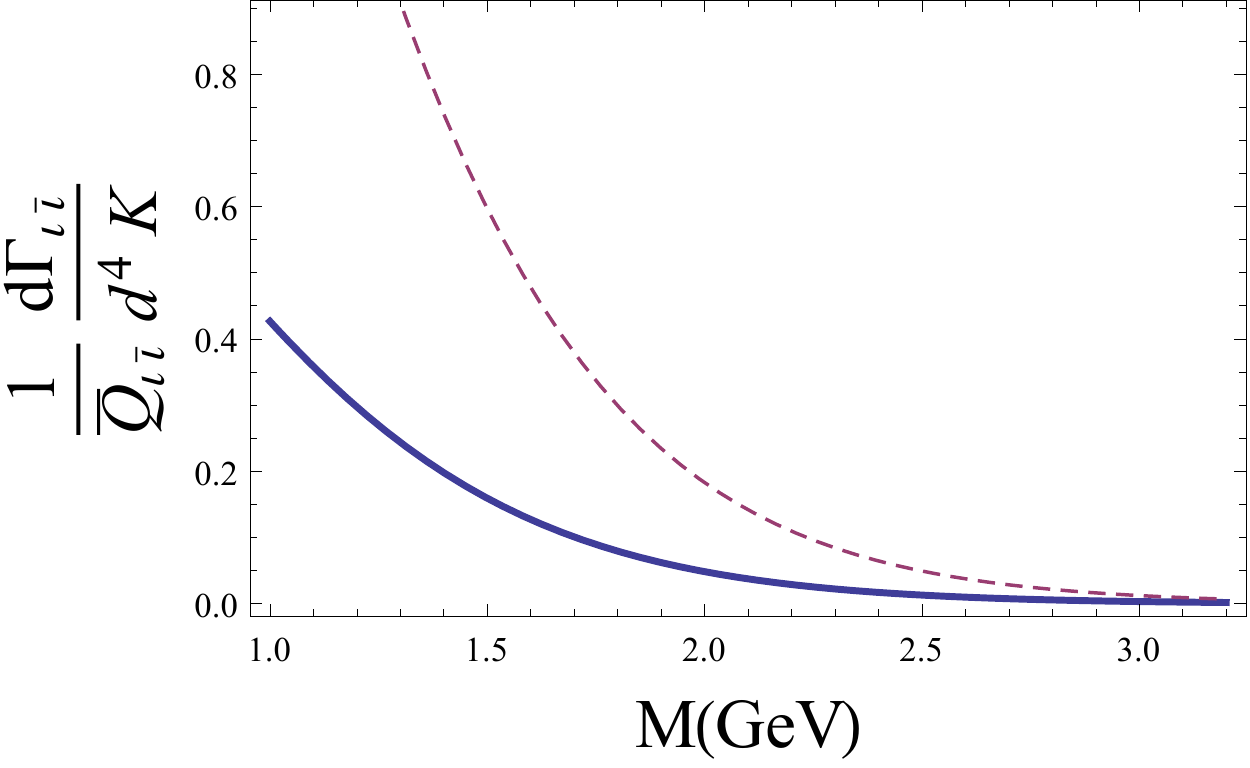}
%\put(-20,42){$T/a$}
%\put(-107,80){$\cb_4/a^4$}
%\put(-107,12){$\cf_4/a^4$}
\\
(\textbf{a}) & (\textbf{b})
\end{tabular}
\caption{Thermal dilepton production in the absence of the magnetic field $B=0$ (\ref{totald1}) [\textit{solid lines}] and in the presence the strong magnetic field $B\gg T^2$ (\ref{totald2}) [\textit{dashed lines}] as a function of the intermediate-mass 1GeV$\leq$ M $\leq$3.2 GeV of the dileptons. In (\textbf{a}) we used $p_{T}=1GeV$, also used (\ref{replacement}) to get $B=B_{SYM}=(6.8\frac{B_{RHIC}}{T^2})\times B_{RHIC}=10.90\times4m_{\pi}^2$, $T=1.58m_{\pi}$, $b=\frac{1}{4\pi^2}\frac{B}{T^2}=0.44$ at \textbf{RHIC}. In (\textbf{b}) we used $p_{T}=1GeV$, also used (\ref{replacement}) to get $B=B_{SYM}=(6.8\frac{B_{LHC}}{T^2})\times B_{LHC}=21.46\times15m_{\pi}^2$, $T=2.18m_{\pi}$ and $b=\frac{1}{4\pi^2}\frac{B}{T^2}=1.72$ at \textbf{LHC}.
\label{dbrl}
}
\end{center}
\end{figure}

Finally, the total thermal dilepton production rates $\frac{d\Gamma^{B=0}_{\ell\bar\ell(Total)}}{d^4 K}$ and $\frac{d\Gamma^{B}_{\ell\bar\ell(Total)}}{d^4 K}$ are
\begin{equation}\label{totald1}
 \frac{d\Gamma^{B=0}_{\ell\bar\ell(Total)}}{d^4 K}\Big|_{\omega\ll T}=2\frac{d\Gamma^{B=0}_{\ell\bar\ell}}{d^4 K}\Big|_{\omega\ll T}=2(-2\frac{p^2_{T}}{\omega^2}+6)Q_{\ell\bar\ell}\omega\sigma(1),
\end{equation}
and
\begin{equation}\label{totald2}
 \frac{d\Gamma^{B}_{\ell\bar\ell(Total)}}{d^4 K}\Big|_{\omega\ll \sqrt{B}}= \frac{d\Gamma^{B_{\parallel}}_{\ell\bar\ell}}{d^4 K}\Big|_{\omega\ll \sqrt{B}}+\frac{d\Gamma^{B_{\perp}}_{\ell\bar\ell}}{d^4 K}\Big|_{\omega\ll \sqrt{B}}=(-4(b+\frac{1}{\sqrt{3}})\frac{p^2_{T}}{\omega^2}+8b+\frac{16}{\sqrt{3}})Q_{\ell\bar\ell}\omega\sigma(1),
\end{equation}
where we used $p_{T}^2=k_{x}^2+k_{z}^2$. We've plotted the total soft-thermal dilepton production rates (\ref{totald1}) and (\ref{totald2}) in Fig.~\ref{dbrl}. Also, note that in Fig.~\ref{dbrl} $\overline{Q}_{\ell\bar\ell}=\frac{Q_{\ell\bar\ell}\sigma(1)T}{n_{b}(k^0)}$, and $\omega^2=p_{T}^2+M^2$.

%The enhancement of the total dilepton production rate can also be quantified by taking the ratio
%\begin{equation}
 % x_{\ell\overline{\ell}}=\frac{\frac{d\Gamma^{B}_{\ell\bar\ell(Total)}}{d^4 K}\Big|_{\omega\ll \sqrt{B}}}{\frac{d\Gamma^{B=0}_{\ell\bar\ell(Total)}}{d^4  %K}\Big|_{\omega\ll T}}=\frac{(\frac{4}{\sqrt{3}}+2b)(\frac{M}{p_{T}})^2+(\frac{4-\sqrt{3}}{\sqrt{3}}+2b)}{3(\frac{M}{p_{T}})^2+2}.
%\end{equation}

\section{\label{sec:conclusion}Conclusion}
We have calculated the DC conductivity tensor $\sigma^{\mu\nu B}$ of strongly coupled $\mathcal{N}=4$ SYM plasma in the presence of strong magnetic field background $B\gg T^2$. We have shown that the component of the tensor $\sigma^{zz B}$ parallel to the magnetic field $B_{z}=B$ increases linearly with $B$ (\ref{t23}) consistently with the lattice QCD result in \cite{Buividovich:2010tn,Kalaydzhyan}.

Using the DC conductivity tensor $\sigma^{\mu\nu B}$, we have calculated the soft-thermal photon and dilepton production rates of strongly coupled $\mathcal{N}=4$ SYM plasma in the presence of the strong magnetic field $B\gg T^2$. We have found that at $\mathbf{RHIC}$ energy scale the thermal photon production rate is only slightly enhanced due to the strong magnetic field $B$, see Fig.~\ref{pbrl2}, hence the effect of the magnetic field $B$ can be neglected at this energy scale but at $\mathbf{LHC}$ energy scale the thermal photon production rate is significantly enhanced, see Fig.~\ref{pbrl2}, therefore, the effect of the magnetic field $B$ becomes very important at this energy scale.

We have also shown that the thermal or intermediate-mass (1GeV$\leq$ M $\leq$3.2GeV) dilepton production rate, in the presence of the strong magnetic field $B$, is significantly enhanced both at $\mathbf{RHIC}$ and $\mathbf{LHC}$ energy scales, thus the enhancement increases with the decreasing of the invariant mass of the dileptons, see Fig.~\ref{dbrl}, which is in a qualitative agreement with the experimentally observed enhancement at $\mathbf{RHIC}$, see the plots for the peripheral collisions in Fig.8 and Fig.13 of \cite{Manninen:2010yf}.

%Using the DC conductivity tensor $\sigma^{\mu\nu B}$, we've calculated the soft-thermal photon and dilepton production rates of strongly coupled $\mathcal{N}=4$ %SYM plasma in the presence of strong magnetic field background $B\gg T^2$. We've found that, even though, the thermal photon production increases linearly with %$B$ (\ref{total2}), the effect of $B$ on the thermal photon production at RHIC and LHC energies is insignificant, see Fig.~\ref{pbrl2}. %Fig.~\ref{pbrl} and .

%However, we've found that the thermal or intermediate-mass (1GeV$\leq$ M $\leq$3.2 GeV) dilepton production in the presence of $B$ is significantly enhanced %even at RHIC and LHC energies, see Fig.~\ref{dbrl}, and the enhancement increases with the decreasing of the invariant mass of the dileptons in a qualitative %agreement with the experimentally observed enhancement at RHIC \cite{Manninen:2010yf}.

In addition to our strong coupling finding in this paper, the enhancement of thermal photon and dilepton production rates due to the strong magnetic field background has also been found at zero coupling in \cite{Tuchin:2012mf,Tuchin:2010gx}. Therefore, it will be interesting to study the effects of the strong magnetic field $B$ on the thermal photon and dilepton production rates in relativistic hydrodynamics and kinetic theory or thermal perturbative QCD at weak coupling.

\acknowledgments
The author thanks Ho-Ung Yee for stimulating discussions and helpful comments on the draft, A. Lewis Licht, Bo Ling, and Misha Stephanov for discussions and reading the draft.

\appendix

\section{\label{Membrane paradigm}Membrane paradigm}

We write the action for a transversal component $A_{i}(\omega,k_{j},u)$ (where $i,j=x,y,z$ with $i\neq j$ and no summation is assumed over $i$ or $j$) as
 \begin{eqnarray}\label{action}
 S=\int d^{5}x\mathcal{L}&=&-\f{1}{2g^2_{5}}\int d^{5}x\sqrt{-g}[g^{uu}g^{ii}\partial_{u}A_{i}(t,k_{j},u)\partial_{u}A_{i}(t,k_{j},u)+g^{tt}g^{ii}\partial_{t}A_{i}(t,k_{j},u)\partial_{t}A_{i}(t,k_{j},u) \nonumber \\
 &+&g^{jj}g^{ii}\partial_{jj}A_{i}(t,k_{j},u)\partial_{j}A_{i}(t,k_{j},u)],
 \end{eqnarray}
where $g^{2}_{5}=\frac{16\pi^{2}R}{N^2}$. After integrating by parts, using the equation of motion and applying Dirichlet's boundary condition at $u=0$ for $A_{i}(\omega,k_{j},u)$, the on-shell action at $u=\epsilon$, becomes
\begin{equation}\label{baction}
S_{on-shell}=-S_{B}[\epsilon],
\end{equation}
where
\begin{equation}\label{baction}
S_{B}[\epsilon]=-\frac{1}{2g_5^{2}}\int_{u=\epsilon}d^{4}x\sqrt{-g}g^{uu}g^{ii}A_{i}(t,k_{j},u)\partial_{u}A_{i}(t,k_{j},u)\ .
\end{equation}
Note that, according to the membrane paradigm \cite{Kovtun:2003wp, Iqbal:2008by}, $S_{B}$ is defined in such away that
\begin{equation}
  S=S_{on-shell}+S_{B}=0.
\end{equation}

We can also calculate the corresponding current at the boundary or any other cut-off hypersurface as
 \begin{equation}\label{current}
 J^{i}=\frac{\partial\mathcal{L}}{\partial \partial_{u}A_{i}}=\frac{\delta S_{B}}{\delta A_{i}}=-\frac{1}{g^{2}_{5}}\sqrt{-g}g^{uu}g^{ii}\partial_{u}A_{i}(t,k_{j},u),
 \end{equation}
by imposing the boundary condition that the conjugate momentum $\Pi^{ui}=\frac{\partial\mathcal{L}}{\partial \partial_{u}A_{i}}$ in the bulk should be equal to the current $J^{i}=\frac{\delta S_{B}}{\delta A_{i}}$ at the boundary or at any other hypersurface at $u=\epsilon$.

According to the membrane paradigm \cite{Kovtun:2003wp, Iqbal:2008by}, in order to evaluate the DC conductivity $\sigma^{ii}$, it's enough to evaluate the current $ J^{i}$ at the horizon $u=1$ where we can use the Eddington-Finklestein coordinate $v$ defined by
\begin{equation}
  dv=dt-\sqrt{\frac{g_{uu}}{-g_{tt}}}du=0,
\end{equation}
at the horizon $u=1$, to re-write
\begin{equation}
\partial_{u}A_{i}(t,z,u)=\sqrt{\frac{g_{uu}}{-g_{tt}}}\partial_{t}A_{i}(t,k_{j},u).
\end{equation}
Therefore, the current (\ref{current}) at the horizon $u=1$ becomes
 \begin{equation}\label{currenth}
  J^{i}=-\frac{1}{g^{2}_{5}}\sqrt{g_{xx}(1)g_{yy}(1)g_{zz}(1)}g^{ii}(1)\partial_{t}A_{i}(t,k_{j},u=1).
 \end{equation}
Remember that we can write the current (or the response for a change in $A_{i}(t,k_{j},u=1)$) in terms of the conductivity $\sigma^{ii}$ as
 \begin{equation}\label{ohms}
  J^{i}=-\sigma^{ii}(u=1)\partial_{t}A_{i}(t,k_{j},u=1),
 \end{equation}
which is nothing but Ohm's law if we recall that the electric field $E_{i}$ is given by $E_{i}=F_{it}=-\partial_{t}A_{i}$, we can compare (\ref{currenth}) with (\ref{ohms}) in order to infer that
\begin{equation}
  \sigma^{ii}(1)=\frac{1}{g^{2}_{5}}\sqrt{g_{xx}(1)g_{yy}(1)g_{zz}(1)}g^{ii}(1).
\end{equation}
 This is exactly what we got in the main text (\ref{txy}) using the RG flow approach.

\section{\label{Equations of motion}Equations of motion}

\subsection{Equation of motion for $B=0$}

$A_{x}$ satisfies the equation of motion (\ref{eom11}) which using the $AdS_{5}$ metric (\ref{ads5}) and a light like momenta, i.e., $k_{z}^2=\omega^2$ can be written as
\begin{eqnarray}\label{eom000}
A''_x + \frac{f'}{f}A'_x +\frac{1-f}{uf^2}\Big(\frac{\omega}{2\pi T}\Big)^2 A_{x}&=&0,
%\p_u\bigg[\f{2\pi^2T^2Rf}{g^2_{5}} \f{A'_{z}}{(1-\frac{k_{z}^2}{\omega^2}f)} \bigg]+\f{RA_{z}\omega^2}{2ufg^2_{5}}&=&0.
\end{eqnarray}
It's easy to see from the above form of the equation of motion that the low frequency limit is achieved for $\omega\ll T$.

\subsection{Equations of motion for $B\gg T^2$}

\paragraph{$\mathbf{k_{z}}\parallel \mathbf{B_{z}}$}
 $A_{x}$ satisfies the equation of motion (\ref{eom11}) which using the $AdS_{3}$ metric (\ref{ads3c}) and a light like momenta, i.e., $k_{z}^2=\omega^2$ can be written as
\begin{eqnarray}\label{eomx}
A''_x + \frac{f_{B}'}{f_{B}}A'_x +\frac{1-f_{B}}{4uf_{B}^2}\Big(\frac{\omega}{2\pi T}\Big)^2 A_{x}&=&0,
%\p_u\bigg[\f{2\pi^2T^2Rf}{g^2_{5}} \f{A'_{z}}{(1-\frac{k_{z}^2}{\omega^2}f)} \bigg]+\f{RA_{z}\omega^2}{2ufg^2_{5}}&=&0.
\end{eqnarray}

\paragraph{$\mathbf{k_{x}}\perp \mathbf{B_{z}}$}

$A_{y}$ satisfies the equation of motion (\ref{eom11}) which using the $AdS_{3}$ metric (\ref{ads3c}) and a light like momenta, i.e., $k_{x}^2=\omega^2$ can be written as
\begin{eqnarray}\label{eomy}
A''_y + \frac{f_{B}'}{f_{B}}A'_y +\frac{\frac{u\sqrt{3}B}{(2\pi T)^2}-f_{B}}{4u^2f_{B}^2}\Big(\frac{\omega}{2\pi\sqrt{\sqrt{3}B}}\Big)^2 A_{y}&=&0,
%\p_u\bigg[\f{2\pi^2T^2Rf}{g^2_{5}} \f{A'_{z}}{(1-\frac{k_{z}^2}{\omega^2}f)} \bigg]+\f{RA_{z}\omega^2}{2ufg^2_{5}}&=&0.
\end{eqnarray}
And $A_{z}$ satisfies the equation of motion (\ref{eom11}) which using the $AdS_{3}$ metric (\ref{ads3c}) and a light like momenta, i.e., $k_{x}^2=\omega^2$ can be written as
\begin{eqnarray}\label{eomz}
A''_z + \frac{(f_{B}u)'}{f_{B}u}A'_z +\frac{\frac{u\sqrt{3}B}{(2\pi T)^2}-f_{B}}{4u^2f_{B}^2}\Big(\frac{\omega}{2\pi\sqrt{\sqrt{3}B}}\Big)^2A_{z}&=&0,
%\p_u\bigg[\f{2\pi^2T^2Rf}{g^2_{5}} \f{A'_{z}}{(1-\frac{k_{z}^2}{\omega^2}f)} \bigg]+\f{RA_{z}\omega^2}{2ufg^2_{5}}&=&0.
\end{eqnarray}
Therefore, we notice from the form of the above equation of motions (\ref{eomy}) and (\ref{eomz}), in the presence of strong external magnetic field $B\gg T^2$, the low frequency limit is achieved for $\omega\ll \sqrt{\sqrt{3}B}$, and all of our results in the main text should be compared to the experiments in this region of the frequency.

\bibliographystyle{JHEP}
\bibliography{ref}

\end{document}